\documentclass[traditabstract]{aa}

\usepackage{graphicx,txfonts,natbib}
\usepackage{epstopdf}
\bibpunct{(}{)}{;}{a}{}{,} 

\newcommand{\gsim}{\;\rlap{\lower 2.5pt
 \hbox{$\sim$}}\raise 1.5pt\hbox{$>$}\;}
\newcommand{\lsim}{\;\rlap{\lower 2.5pt
   \hbox{$\sim$}}\raise 1.5pt\hbox{$<$}\;}
\newcommand{\be}{\begin{equation}}
\newcommand{\beq}{\begin{equation}}
\newcommand{\ba}{\begin{eqnarray}}
\newcommand{\ee}{\end{equation}}
\newcommand{\eeq}{\end{equation}}
\newcommand{\ea}{\end{eqnarray}}

\newcommand{\bea}{\begin{eqnarray}}
\newcommand{\eea}{\end{eqnarray}}
\newcommand{\bean}{\begin{eqnarray*}}
\newcommand{\eean}{\end{eqnarray*}}

\newcommand{\bx}{{\bf x}}

\newcommand{\bk}{{\bf k}}
\newcommand{\bu}{{\bf u}}
\newcommand{\bn}{{\bf \hat{n}}}

\newcommand{\tcmb}{T_\gamma}
\newcommand{\Lyman}{${\rm Ly\alpha}$ }

\def\msun{{\rm\,M_\odot}}
\newcommand{\lya}{${\rm Ly\alpha}$ }

\newcommand{\meanxa}{$\langle x_\alpha\rangle$ }

\begin{document}

\title{Probing the first galaxies with the SKA} 
\author{M.G.~Santos\inst{1} \and M.B.~Silva\inst{1} \and J.R.~Pritchard\inst{2}\thanks{Hubble Fellow} \and R. Cen\inst{3} \and A. Cooray\inst{4}}
\institute{
  CENTRA, Departamento de F\'isica, Instituto Superior T\'ecnico, 1049-001 Lisboa, Portugal\\
  email: mgrsantos@ist.utl.pt
  \and
  Harvard-Smithsonian Center for Astrophysics, Cambridge, MA 02138, USA
  \and
  Department of Astrophysical Sciences, Princeton University, Princeton, NJ 08544, USA
  \and
  Center for Cosmology, Department of Physics and Astronomy, University of California, Irvine, CA 92697, USA
}

\date{Received 5 September 2010 / Accepted 26 December 2010}

\abstract{
Observations of anisotropies in the brightness temperature of the 21 cm line of neutral hydrogen from the period before reionization would shed light on the dawn of the first stars and galaxies.  In this paper, we use large-scale semi-numerical simulations to analyse the imprint on the 21 cm signal of spatial fluctuations in the Lyman-$\alpha$ flux arising from the clustering of the first galaxies.  
We show that an experiment such as the Square Kilometer Array (SKA) can probe this signal at the onset of reionization, giving us important information about the UV emission spectra of the first stars and characterizing their host galaxies. SKA-pathfinders with $\sim10\%$ of the full collecting area should be capable of making a statistical detection of the 21~cm power spectrum at redshifts $z\lesssim20$ (corresponding to frequencies $\nu\gtrsim 67$ MHz). We then show that the SKA should be able to measure the three dimensional power spectrum as a function of the angle with the line of sight and discuss the use of the redshift space distortions as a way to separate out the different components of the 21~cm power spectrum. We demonstrate that, at least on large scales where the \lya fluctuations are linear, they can be used as a model independent way to extract the power spectra due to these \lya fluctuations.
}{}{}{}{}


\maketitle

\section{Introduction}

The period of formation of the very first stars is one of the least understood epochs in the history of the Universe.
The 21~cm spin-flip transition of neutral hydrogen at high redshifts has the potential to open a new observational window to study this early period, 
where the most distant, first galaxies reside,
and even beyond that. 
Moreover, 21~cm observations
may provide the only means to detect signatures
of the first galaxies at $z>15$ in the foreseeable future.
While the {\em James Webb Space Telescope} (JWST\footnote{http://www.jwst.nasa.gov/}) will be able to detect galaxies
at $z\le 10$ \citep[e.g.,][]{2004Stiavelli}, it will not image all of the sources responsible for the reionization of the Universe.
If current estimates in the literature are correct, even deep fields with JWST will only detect $\sim 30\%$ of the sources responsible for reionization
or maintaining reionization at $z \sim 12$ \citep{salvaterra10}. At $z \sim 20$, direct emission from stars will be below the threshold of
JWST observations and new techniques will be required to probe the onset of reionization.
One possibility is Lyman-$\alpha$ emitter surveys over the next decade, but even these surveys will only be able to detect the brightest
galaxies for selected redshift ranges at $z\le 10$
\citep[e.g.,][]{2008Ouchi,
2007Nilsson,
2007Cuby,
2007Stark,
2008Willis,
2008McMahon,
2009Hibon},
but not the fainter, first galaxies at redshift $z\ge 20$.
The next generation of large ground based telescopes, like the {\em European Extremely Large Telescope} (E-ELT\footnote{http://www.eelt.org}), {\em Thirty Meter Telescope} (TMT\footnote{http://www.tmt.org/}), and {\em Giant Magellan Telescope} (GMT\footnote{http://www.gmto.org/}), will only be capable of detecting at most the very brightest galaxies at redshifts $z\approx20$.  Most of the star formation however will be in fainter galaxies making it impossible for these telescopes alone to survey the full population.  In combination with 21 cm observations, which are sensitive to the integrated \lya emission of the galaxies, it should be possible to form a complete census of the first sources.

With a wide frequency coverage,
observations of the 21~cm signal will provide a 3-dimensional 
tomographic view of the inter-galactic medium (IGM), 
before and during the cosmological reionization \citep[for a review see][]{furlanetto06c}.
As soon as the first galaxies appear, 
at a redshift of $z\sim 20-30$ in the standard cold dark matter cosmological model
\citep[e.g.,][]{komatsu2010}, photons emitted at frequencies between the \Lyman and Lyman limit 
quickly couple the spin temperature of the hyperfine level population to the 
IGM gas temperature through the Wouthuysen-Field effect \citep{wouth1952,field1959spint}. 
Provided the IGM has been cooling adiabatically (X-ray heating is still negligible during this early stage of galaxy formation) ,
the 21~cm signal will be strong and observed in absorption against the cosmic microwave background (CMB). 
Because the \Lyman coupling depends on the local \Lyman radiation, 
which is correlated with galaxies,
the 21~cm signal fluctuations will trace the location, mass, emission spectra, 
luminosity and redshift evolution of high redshift galaxies. 

Since the first galaxies represent rare high-$\sigma$ peaks in the cosmic density field
\citep[e.g.,][]{2007Trac},
their spatial distribution have large fluctuations and we expect the \Lyman contribution 
to the overall 21~cm signal to be appreciable, 
considerably improving the prospects for detection of the \Lyman fluctuations through 21~cm observations.
Although the high redshift-end of the reionization process 
will be inaccessible to the first generation experiments, such as the Low Frequency Array (LOFAR\footnote{http://www.lofar.org}) and the Murchison Widefield Array (MWA\footnote{http://www.mwatelescope.org}),
the second generation experiments such as the Square Kilometre Array
(SKA\footnote{http://www.skatelescope.org}) should have enough collecting area to statistically probe the signal.

On the theoretical front, analytical models have been useful in predicting the probable evolution of the high redshift signal \citep{barkana05b,pritchard06,pritchard2008}, but rely on linear approximations and require further comparison with simulations. Numerical simulations, on the other hand, can provide a self-consistent treatment of the \Lyman radiative transfer, taking into account effects such as the scattering in the \Lyman line wings \citep{semelin2007, baek09}, although they are typically slow to run and are limited to small volumes ($\leq 100$ Mpc/h). Recent developments using semi-numerical algorithms have allowed the rapid generation of the high redshift 21cm signal during the pre-reionization epoch \citep{santos08,santos10,mesinger10} in large boxes, while maintaining the 3-d structure of the signal on large scales as seen in the full numerical simulations. 

In this paper we will generate fast simulations as described in \citet{santos10} to analyze the dependence of the overall 21~cm 
signal on several parameters related to the first galaxies in the Universe and consider the ability of an experiment like SKA to constrain these parameters. 
The layout of this paper is as follows. We start in Section~2 by describing the 21~cm signal and the free parameters of the model that affect the \Lyman fluctuations. In 
Section~3 we present the experimental setup used for SKA, calculating the expected error on the 3-d power spectrum. 
In Section~4 we study the possibility of constraining the signal in a model independent way, using the redshift-space distortions and the corresponding measurements on the 3-d power spectrum as a function of the angle with respect to the line of sight.  We end with our conclusions and a discussion of the prospects for SKA in Section~5.

Throughout this paper where cosmological parameters are required we use the standard set of values $\Omega_m=0.28$, $\Omega_\Lambda=0.72$, $\Omega_b=0.046$, $H=100h\,\rm{km\,s^{-1}\,Mpc^{-1}}$ (with $h=0.70$), $n_S=0.96$, and $\sigma_8=0.82$, consistent with the latest measurements \citep{komatsu2010}.

\section{The 21~cm signal: \Lyman fluctuations}

The 21~cm brightness temperature corresponds to the change in the
intensity of the CMB radiation due to absorption or emission when it
travels through a patch of neutral hydrogen.  It is given, at an
observed frequency $\nu$ in the direction $\bn$, by (see e.g. \citealt{santos08})
\begin{eqnarray}
\label{t21}
&&\delta T_b(\nu) \approx  23 x_{HI}(1+\delta)
\left(1-\frac{T_{\gamma}}{T_S}\right)
\left( \frac{h}{0.7}\right)^{-1}
\left( \frac{\Omega_b h^2}{0.02} \right)\times\nonumber\\ 
&&\left[\left(\frac{0.15}{\Omega_m h^2} \right)\left(\frac{1+z}{10}\right)
\right]^{1/2}\left(\frac{1}{1+1/H\,dv_r/dr}\right)\ \rm{mK},
\end{eqnarray}  
where $x_{\rm HI}$ is the fraction of neutral hydrogen (mass weighted), $dv_r/dr$ is the comoving gradient of the line of sight component of the comoving velocity and we use $\delta_a$ for the fractional value of the quantity $a$ ($\delta_a\equiv {a-\langle a\rangle\over \langle a\rangle}$) with $\delta$ for the fluctuation in the matter density.

The spin temperature ($T_S$) is coupled to the hydrogen gas temperature
($T_K$) through the spin-flip transition, which can be excited by
collisions or by the absorption of \Lyman photons (Wouthuysen-Field
effect) and we can write: 
\be 1-{\tcmb\over T_S}={x_{tot}\over
  1+x_{tot}}\left(1-{\tcmb\over T_K}\right), 
\ee where
$x_{tot}=x_\alpha+x_c$ is the sum of the radiative and collisional
coupling parameters and we are already assuming that the color
temperature of the \Lyman radiation field at the \Lyman frequency is
equal to $T_K$. Note that although we talk here about \Lyman radiation, in effect we consider the contribution from all the photons up to the Lyman limit frequency, since photons redshifting into Lyman series resonances can produce \lya photons as a result of atomic cascades \citep{hirata2006lya,pritchard2006}. When the coupling to the gas temperature is
negligible (e.g. $x_{tot}\sim 0$), $T_S\sim \tcmb$ and there is no
signal. On the other hand, for large $x_{tot}$, $T_S$ simply follows
$T_K$.

\subsection{Simulations}

We use the code SimFast21\footnote{Available at http://simfast21.org} 
described in \citet{santos10} to calculate the spin temperature and the corresponding 21~cm signal. 
The code starts by calculating the linear density field in a box 
followed by the corresponding halo and velocity fields.
Then, it computes the nonlinear density and halo fields, 
the collapsed mass distribution, the corresponding star formation rate, the IGM gas temperature, the \Lyman coupling, the collisional coupling and finally the 21~cm 
brightness temperature accounting for all the contributions including the correction due to redshift space distortions using the nonlinear velocity field. 
In order to probe the full range of $k$ space, we generated two end to end simulations using the SimFast21 code, one large simulation with 1Gpc in size and 1.6 Mpc 
resolution ({\bf S1}) and another with higher resolution (0.186 Mpc) but smaller 143 Mpc in size (simulation {\bf S2}).
This code has the advantage of being modular, allowing to replace some of the steps by simulation boxes generated by other techniques. 
Therefore, for comparison, we also applied our \Lyman calculation to the output (star formation rate, matter density) from the \citet{trac08} simulation, thus providing a check of 
the approximation used to generate the halo mass function. 
We call this simulation of the 21~cm signal, {\bf S3}, with a size of 143 Mpc and a resolution of 0.186 Mpc. Table~\ref{table_sims} shows a summary of the simulations used.
\begin{table}
\centering                         
\begin{tabular}{c c c c}        
\hline\hline                 
    & Size & Resolution & Resolution\\
    &      &  (halos)   &  (\lya)\\
\hline
   {\bf S1} & 1000 Mpc & 0.56 Mpc & 1.667 Mpc \\
   {\bf S2} & 143 Mpc & 0.09 Mpc  & 0.186 Mpc \\
   {\bf S3} (N-body) & 143 Mpc & - & 0.186 Mpc \\
\hline                                  
\end{tabular}
\caption{Simulations used in the analysis. Middle column shows the resolution used to resolve halos in the semi-numerical code, while the last column shows the resolution used in the \lya calculation.}
\label{table_sims}     
\end{table}
All simulations use halos with masses down to $10^8 \msun$ corresponding to a minimum virial temperature of $T\sim10^4$~K. Simulation {\bf S1} cannot resolve halos down to these mass scales and we populate the remaining halos in each (empty) cell using a Poisson sampling biased with the underlying density field, giving a mass function consistent with N-body simulations once non-linear corrections are applied, as described in \citet{santos10}. With the new version of the code, we also allow a hybrid approach instead of the ``Poisson'' method, where, after the halo finding algorithm, we calculate the unresolved collapsed mass in each empty cell using the collapsed fraction $f_{\rm coll}$  from the extended  Press-Schechter formula (see e.g. \citealt{zahn07}). This is fine as long as we do not want to resolve halos within a cell. Note that this is used at the cell level, allowing to also apply the non-linear corrections to each cell. Both approaches give similar results in terms of the 21~cm signal, although the later method can be faster as we decrease the halo minimum mass used in the simulation.

Simulation {\bf S3} was run using a hybrid simulation code 
in modeling cosmic reionization \citep{trac08},
incorporating N-body, 
hydrodynamic, and RT algorithms to solve the coupled evolution of the dark matter, baryons, and radiation \citep{2004TracPen, 2006TracPen, 2007Trac}. 
It covers a large dynamic range and satisfies the requirements of having sufficiently high resolution 
to capture small-scale structure, and simultaneously a sufficiently large volume to 
reduce sample variance \citep{2004BarkanaLoeb}. 
The simulation is run in two steps.
The first step involves running a high-resolution N-body simulation with $3072^3$ dark matter
particles on an effective mesh with $11520^3$ cells in a comoving box, $143$Mpc on a side. 
We identified collapsed dark matter halos on the fly using a friends-of-friends algorithm, 
with a linking length $b=0.2$ times the mean inter-particle spacing, in order to model
 radiation sources and sinks. With a particle mass resolution of $2.68\times10^6h^{-1}\msun$,
we can reliably locate all dark matter halos with virial temperatures above the atomic cooling limit 
($T\sim10^4$~K) with a minimum of $\sim40$ particles \citep{2007Heitmann}, 
and half of this collapsed mass budget
is resolved with $>400$ particles per halo. Our halo mass functions are in very good agreement with 
other recent work \citep[e.g.][]{2007ReedBFJT, 2007LukicHHBR, 2008CohnWhite}. 
The second step produces a hydrodynamic + RT simulation with moderate resolution, but incorporating subgrid physics modeled using the high-resolution 
information from the large N-body simulation. 
Radiation sources are prescribed and star formation rates calculated using the halo model described 
in \citet{2007Trac}. 
We consider only Population~II stars from starbursts \citep{2003Schaerer} as contributing to the ionizing photon budget. 
The hydro+RT simulation utilizes equal numbers ($N=1536^3$) of dark matter particles, gas cells, and adaptive rays, where for the latter, we track 5 frequencies above the hydrogen ionizing threshold of 13.6 eV. The photo-ionization and photo-heating rates for each cell are calculated from the incident radiation flux and used in the non-equilibrium solvers for the ionization and energy equations. The initial conditions are generated with a common white noise field and a linear transfer function calculated with CAMB \citep{2000Lewis}.  

In order to calculate the \Lyman coupling in a given cell, the algorithm assumes that the scattering rate is proportional to the total Lyman series flux arriving in that cell (see again \citealt{santos10}). This Lyman series flux is obtained through a 3-dimensional integration of the comoving photon emissivity, $\epsilon(\bx,\nu,z)$ (defined as the number of photons emitted at position $\bx$, redshift $z$ and frequency $\nu$ per comoving volume, per proper time and frequency), which is assumed proportional to the star formation rate:
\be 
\epsilon_\alpha(\bx,\nu, z) = {\rm SFRD}(\bx,z)\epsilon_b(\nu)\ , 
\label{emit2} 
\ee 
where ${\rm SFRD}(\bx,z)$ is the star formation rate density from the simulation (in terms of
the number of baryons in stars per comoving volume and proper time). As stated above, in the simulation S3 we used the star formation rate obtained from the \citet{trac08} simulation. $\epsilon_b(\nu)$ is the spectral distribution function of the
sources (defined as the number of photons per unit frequency emitted
at $\nu$ per baryon in stars).
We assumed a power law model for $\epsilon_b(\nu$):
\be
\epsilon_b(\nu)=A\nu^\alpha,
\ee
between $\nu_\alpha$(10.2 eV) and the Lyman limit frequency (13.6 eV).
This will allow us to calculate how sensitive the 21~cm power spectrum is to
changes in the parameters of the emission model.

We take $\alpha=-0.9$
and $A$ such that the integration between the \lya and Lyman limit frequencies gives a total emission of
20000 photons per baryon. These numbers are based on the expected PopII spectra from \citet{schaerer03}. Note that the parameter $A$ will be totally degenerate with the star formation rate efficiency in our model. 
Since the initial mass function (IMF) is likely to be dominated by massive stars, this single power law model is likely to be a good description of the emission spectra \citep[see e.g.][]{leitherer1999}, although a broken power law can sometimes provide a slightly better fit \citep{pritchard2006}. Moreover, as we shall see later, results are very insensitive to $\alpha$, essentially depending on the total number of photons emitted per baryon.
We also note that the above calculation of \Lyman radiation does not take into account
full radiative transfer effects via their scattering off of neutral hydrogen atoms (scattering in the wings of the \lya line, \citealt{chuzhoy2007,semelin2007}, could increase the amplitude of the fluctuations on small scales).
As a result, we expect that some details of the signal on small scales due to \Lyman coupling
to be partly inaccurate.

Since the details of the galaxy emissivity depend upon the assumed initial mass function (IMF), which is poorly known \citep{barkana2001}, these values are subject to considerable uncertainty.  Constraints would therefore provide useful information about the properties of the first galaxies.  These uncertainties make the mapping between the mean value of the \lya coupling \meanxa and redshift uncertain.  We will therefore describe the shape of the \lya power spectrum for a given \meanxa and caution that the redshift at which that actually occurs may be very different from our fiducial model.

\subsection{Comparison}

\begin{figure}[!t]
\hspace{-0.5cm}
\includegraphics[scale=0.45]{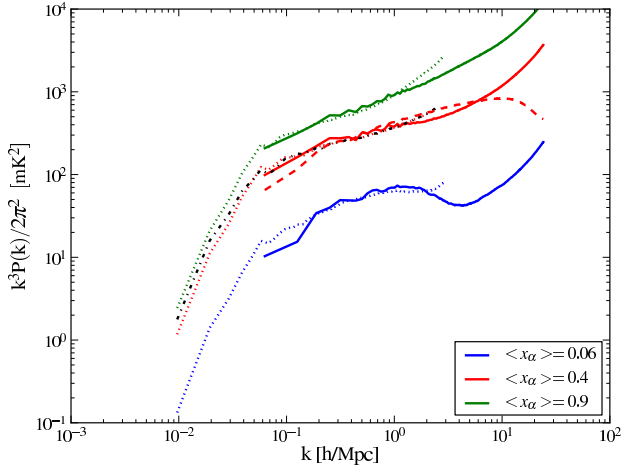}
\caption{The 21~cm brightness temperature with all fluctuations included. Dotted - simulation S1, solid - simulation S2, dashed - simulation S3. Redshifts from top to bottom are: 19.25 ($\langle x_\alpha\rangle = 0.9$), 20.25 ($\langle x_\alpha\rangle = 0.4$), 22.5 ($\langle x_\alpha\rangle = 0.06$). The black dash-dotted line shows the correction for $\langle x_\alpha\rangle = 0.4$ when doing the full \lya flux inhomogeneous integration.}
\label{fig_t21_ps}
\end{figure}

The 21~cm signal occurs over a wide range of scales, with \lya fluctuations contributing at wave-numbers $0.01\lesssim k/(h/{\rm Mpc})\lesssim10$.  Since none of our simulations can cover this broad set of scales individually, we exploit the modularity of the code to piece together two simulations to cover the full range.  To check that this gives self-consistent results, in Figure \ref{fig_t21_ps} we plot the full 21~cm brightness temperature for the three simulations at a few values of the mean \lya coupling $\langle x_\alpha \rangle$.  In each case, we take into account fluctuations from the density, velocity and \Lyman terms as well as the collisional coupling, ionization fraction (although at high redshifts this is negligible) and gas temperature (note again that for simulation S3 we calculate the temperature using our semi-numerical code based on the simulation SFRD). We see a smooth transition from the S1 to the S2 simulation, showing for the first time, the full range of scales from $k\sim 0.01$ h/Mpc to $k\sim 20$ h/Mpc with an amplitude that is roughly between 100 mK$^2$ and 1000 mK$^2$ on the range of interest, which should help the detection of the signal.  The increase in the amplitude as the coupling increases is due primarily to the cooling of the IGM gas in this redshift range, which leads to a stronger absorption signal. 
Note that, as explained in \citet{santos10}, when integrating the \lya flux, we assume the star formation rate is homogeneous above a certain large scale in order to make the calculation faster. In this paper this scale was set to 143 Mpc which is the origin of the slight break we see in figure~\ref{fig_t21_ps} for $x_\alpha \gtrsim 1$. The black dash-dotted line in this figure shows the correction for $x_\alpha=0.4$ if instead we did the inhomogeneous integration up to 300 Mpc. Since this is a small correction, we decided to keep the limit to 143 Mpc.

For comparison, the dashed line shows the power spectrum from simulation S3, which agrees quite well with results using the semi-numerical code validating our halo/star formation prescription (although in S3 there is a decrease of the signal on very small scales, which is related to a smoothing of the velocity term on these scales). An extensive comparison made by \citet{baek10} using radiative transfer simulations (although for smaller box size $\sim 100$ Mpc/h and lower mass resolution $\gtrsim 10^{10} M_\odot$) shows a similar spectrum for models in the same parameter range.

\begin{figure}[!t]
\includegraphics[scale=0.45]{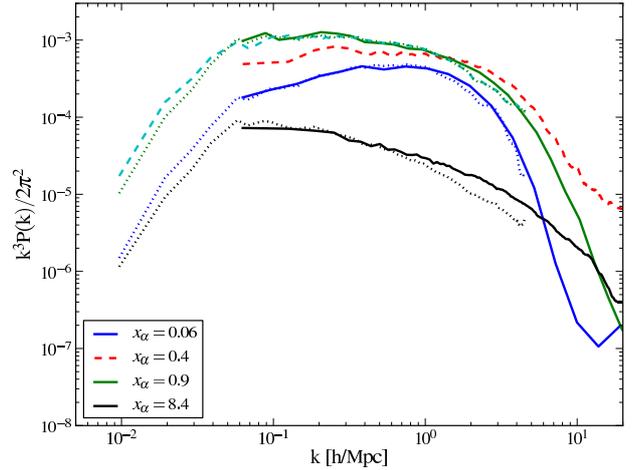}
\caption{Power spectrum of ${x_\alpha \over 1 + x_\alpha}$ for several values of $\langle x_\alpha\rangle$. Dotted lines correspond to simulation S1, solid lines to simulation S2 and the top dashed line to simulation S3. Redshifts from top to bottom are: 19.25 ($\langle x_\alpha\rangle = 0.9$), 20.25 ($\langle x_\alpha\rangle = 0.4$), 22.5 ($\langle x_\alpha\rangle = 0.06$), 18.75 ($\langle x_\alpha\rangle = 8.4$). Again, the top, cyan dashed line shows the correction when doing the full calculation.}
\label{fig_xacoup}
\end{figure}

Since it will be the focus of the present paper, we now examine the \lya fluctuations in more detail.  
In Figure~\ref{fig_xacoup} we show the power spectrum of the quantity $A_{\alpha}\equiv{x_\alpha\over 1+x_\alpha}$ 
for the three simulations at four different values of $\langle x_\alpha \rangle$. 
$A_{\alpha}$ represents the contribution of the \Lyman coupling to the 21~cm signal
and varies between 0 and 1.
Note that the $x_\alpha$ field tends to be dominated by the large values close to individual sources, which obscures the details of the coupling since in these regions $A_{\alpha}$ saturates. 
The amplitude of the power spectrum principally depends on $\langle x_\alpha \rangle$ with the fluctuations peaking at $\langle x_\alpha\rangle\sim 0.5$,
suggesting that this is the regime to best extract information related to the Wouthuysen-Field effect ($z\sim 20$ in our model),
and we will concentrate on these redshifts from now on. 
Figure~\ref{fig_xacoup} shows again the correction to the calculation if we do not impose a limit to the \lya inhomogeneous integration (top, cyan dashed line, for $x_\alpha=0.9$). This correction will be more evident as $x_\alpha$ increases (see the "break" for  $x_\alpha=8.4$) but at the same time the contribution to the overall 21cm signal will become smaller so that it should be safe to neglect this correction.

Comparing to figure~14 in \citet{baek09} we see that the fluctuation level is similar, 
although we do not see their small scale increase in power due to point sources for $k>1 {\rm Mpc/h}$.
The difference on small scales ($k>1 {\rm Mpc/h}$) between our results and those of 
\citep{baek09} is, at least in part, likely due to lack of \Lyman radiative transfer treatment in ours (although differences in size and mass resolution of the simulations should also be considered). As will become clear later, it is the 21~cm observations at large scales that offer the best means to extract \Lyman physics,
so the inaccuracies at small scales do not pose a significant problem.

The full N-body simulation (S3) agrees reasonable well with our calculation
and again we see a smooth transition from the very large scale simulation to the high resolution one, giving us a complete view over all scales of the expected signal.  Although it is clear that there is room for more detailed numerical work on the 21 cm signal, these comparisons give us confidence in the semi-analytic simulation and we now turn to exploring the consequences for observations.

\subsection{Identifying the \Lyman epoch}

\begin{figure}[!t]
\hspace{-0.5cm}
\includegraphics[scale=0.45]{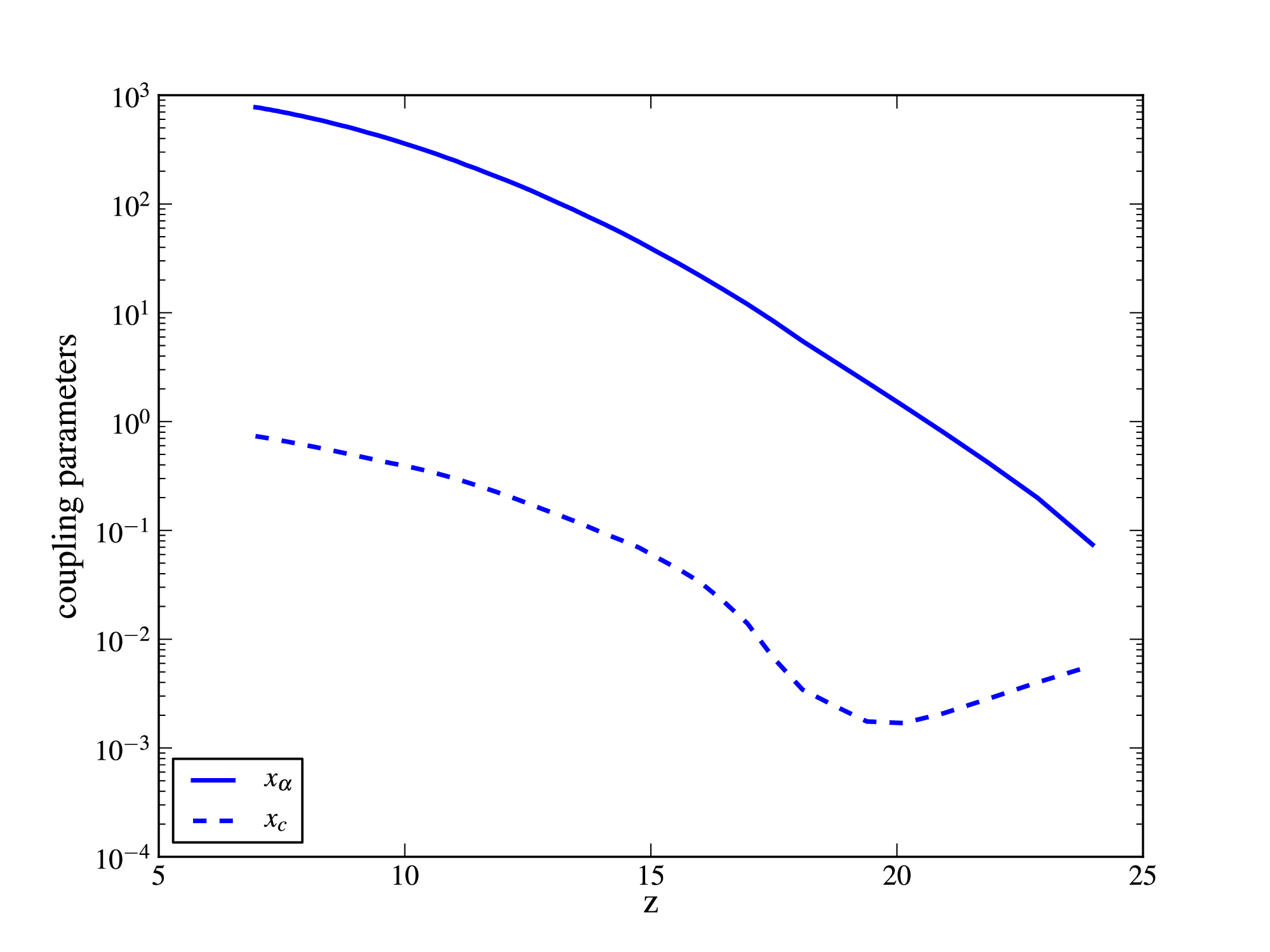}
\caption{The average $x_\alpha$ and $x_c$ coupling parameters as a function of redshift for the S2 simulation. The 
\Lyman coupling dominates over collisions at all redshifts considered.}
\label{fig_xa_xc}
\end{figure}

\begin{figure}[!t]
\includegraphics[scale=0.45]{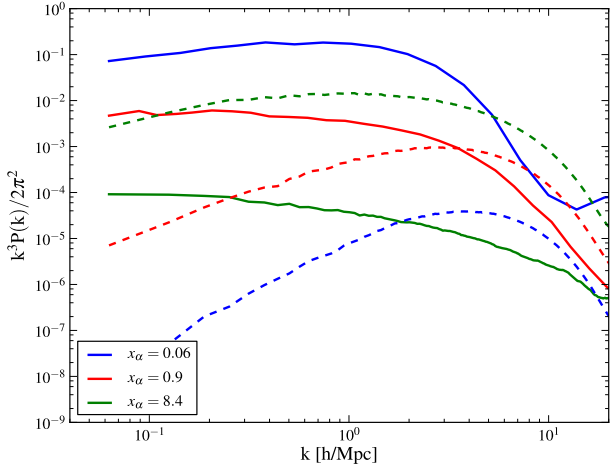}
\caption{The fluctuations from \Lyman, $A_{\alpha}$ (solid) and the gas temperature, $1-{\tcmb\over T_K}$ (dashed), contributing to the 21~cm brightness temperature.
Note that both quantities are normalized and the gas temperature fluctuations only include the x-ray heating.}
\label{fig_xa_TK}
\end{figure}

In principle, there are several contributions to the 21cm brightness temperature and full modeling of the signal would be required to disentangle the part due to \lya fluctuations and obtain parameter constraints on the first galaxies. However, as we can see from figure~\ref{fig_xa_xc} and figure~\ref{fig_xa_TK}, both the fluctuations in the collisional coupling and in the gas temperature should be subdominant at the redshifts where the \Lyman fluctuations are strongest (when $\langle x_\alpha \rangle < 1$ and $k< 1 h/{\rm Mpc}$) making the analysis of the signal more straightforward (the ionization is very small at these redshifts, so that their contribution to the fluctuation power is completely negligible). The fluctuations in the gas temperature shown in figure~\ref{fig_xa_TK} only include the heating due to x-rays, which can potentially be harder to separate from the \lya fluctuations. On top of this we also have fluctuations in the gas temperature due to the adiabatic cooling process which are essentially proportional to the matter density perturbations. Both contributions are intrinsically included in the simulation, although, as we see, the effect from x-ray heating is negligible.


The question then is if whether we can actually identify this epoch from observations rather than theoretical guesses. In many models, the global (average) 21~cm signal clearly shows the transition from \Lyman coupling to X-ray heating and finally to emission once the gas is heated above the CMB temperature \citep{pritchard10}. This average signal will not be accessible to interferometers, but as was shown in \citet{santos08}, figure 9, the large scale 21~cm power spectrum can also depict the evolution of the signal. 

\begin{figure}[!t]
\includegraphics[scale=0.45]{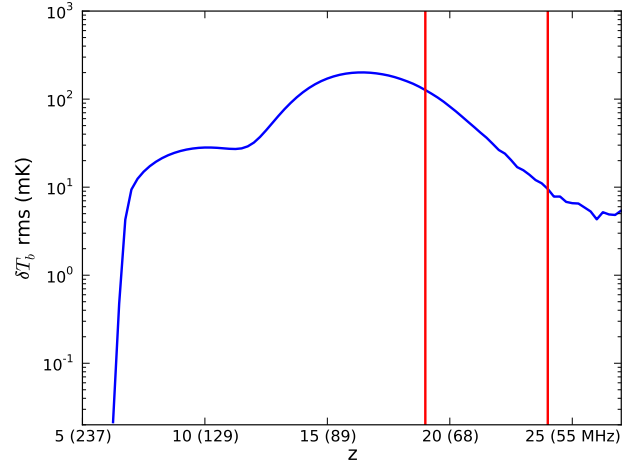}
\caption{$\sqrt{\langle \left(\delta T_b-\langle \delta T_b \rangle\right)^2\rangle}$ (rms) of the signal as a function of redshift for simulation S2 (frequency in parenthesis). Red lines indicate the region where we can safely ignore fluctuations from collisions and X-ray heating.}
\label{t21rms}
\end{figure}

The rms of the signal ($\langle \left(\delta T_b-\langle \delta T_b \rangle\right)^2\rangle$) can also provide that information and should be easier to measure with interferometers. As we can see in figure \ref{t21rms}, at $z>25$ the signal is initially small since coupling to the gas temperature is low. As the \Lyman coupling increases, the spin temperature approaches the gas temperature and the rms of the signal increases, not only due to fluctuations on the \Lyman field but also because the factor $A_{\alpha} \left(1-T_\gamma/T_K\right)$ is increasing (in absolute terms) since the gas is still cooling adiabatically. Once the gas temperature begins to increase due to X-ray heating the signal decreases and we see the turning point at $z\sim 16$ (the transition from absorption to emission occurs later at $z\sim 13$.  Finally, the gas is heated far above the CMB temperature and the signal plateaus until reionization finally causes it to die away at $z\sim7$.

Note that as X-rays start heating the IGM, the signal does not drop immediately since this heating is inhomogeneous and there will be some region still cooling adiabatically as well as temperature fluctuations contributing to the rms. Therefore we see a small plateau at the peak of the signal starting at $z\sim 19$. The conclusion is that it should be safe to neglect the contributions of the X-ray heating to the 21~cm signal between the points when the rms (or the average) starts rising at $z\gtrsim 24$ and before we reach the plateau at the maximum $z\lesssim 19$ ($x_\alpha \sim 0.9)$, at least on large scales where the \Lyman contribution dominates over the temperature as we can see in figure~\ref{fig_xa_TK}.

\begin{figure}[!t]
\hspace{-0.5cm}
\includegraphics[scale=0.45]{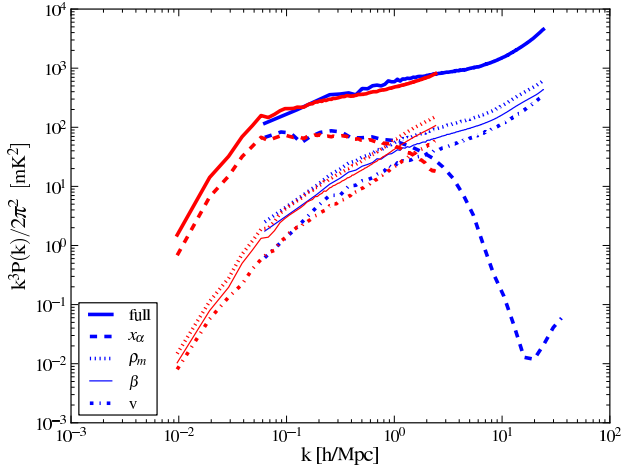}
\caption{The 21~cm power spectrum and its main contributions at $\langle x_\alpha \rangle=0.4$ ($z\sim 20.25$) using simulation S1 (red) and simulation S2 (blue). Top solid lines - all contributions included; dashed - \lya only; dotted - matter density; bottom thin solid lines - fluctuations on the gas temperature from adiabatic cooling only; dot-dashed - velocity fluctuations only.}
\label{fig_contrib}
\end{figure}

In this case, the equation simplifies and we can write the full 21~cm signal as:
\be
\delta T_b(\nu) = C(z) (1+\delta)(1+\delta_{A_{\alpha}})(1+\delta_{A_v})(1+\beta\delta),
\label{t21_d}
\ee
where $\beta=2/3\bar{T}_K/(\bar{T}_K-T_\gamma)$ (note the different definition from \citealt{barkana05b}), corresponding to the adiabatic cooling process ($T_K\propto \rho^{2/3}$) and $\bar{T}_K\approx 180\left(\frac{1+z}{100}\right)^2$ K.
Also,
\bea
&&C(z) \approx  23 \left( \frac{0.7}{h}\right)\left( \frac{\Omega_b h^2}{0.02} \right)
\left[\left(\frac{0.15}{\Omega_m h^2} \right)\left(\frac{1+z}{10}\right)\right]^{1/2}\times \\ \nonumber
&& \langle 1-\frac{T_{\gamma}}{T_K} \rangle \langle A_{\alpha} \rangle \langle A_v \rangle\ \ \rm{mK}
\eea
with $A_v=\left(\frac{1}{1+1/H\,dv_r/dr}\right)$ ($\langle A_v \rangle \sim 1$).
Figure~\ref{fig_contrib} shows the contribution to the full 21~cm signal from each of the terms considered on equation \ref{t21_d} using simulations S1 and S2. Again we see that the \Lyman term dominates over large scales $k<1\ $h/Mpc, which makes it all important to be able to generate large volume simulations ($>100$ Mpc/h) and give us confidence that it should be possible to extract the \Lyman signal over the redshift range proposed without confusion from other contributions. Moreover, the small contribution at these scales from the other terms in equation~\ref{t21_d} can be modeled with reasonable accuracy if we know the matter power spectrum.

\section{Measurements with the SKA}

In this Section, we outline details related to the experimental measurement of the 21~cm signal at the very high redshifts during the epoch where \lya dominates.

\subsection{Noise Power Spectrum}

Before we go into the details of the experimental setup, we quickly review the noise calculation that goes into our analysis. Following \citet{bowman2006,mao08,mcquinn06}, the 
expected error $\Delta P(k,\theta)$ on the measurement of the 3-d power spectrum of the signal $P_S(k,\theta)$ with noise $P_N(k,\theta)$ is given by
\be
\Delta P(k,\theta) = \frac{1}{\sqrt{N_m(k,\theta)}}\left[P_S(k,\theta)+P_N(k,\theta)\right]\,,
\label{pk_error}
\ee
where we are assuming that the power spectrum depends only on the moduli ($k$) of the vector $\bk$ and on the angle $\theta$ between $\bk$ and the line of sight. $N_m(k,\theta)$ is the total number of modes in $\bk$ space contributing to the measurement (note that the sum is only done over half the sphere). In order to obtain the above expression, we ``grid'' the $\bk$ space into pixels of size $dk_\perp^2 dk_\parallel$, where $\perp$ denotes the component perpendicular to the line of sight and $\parallel$ the parallel one and assume that measurements in different pixels are uncorrelated. The size of these pixels is just determined by the real space volume of the observed region: 
\be
{\rm Vol}=r^2 y A_{FoV} B\,,
\ee
where $A_{FoV}$ is the field of view of the experiment (in radians), $B$ is the bandwidth for this particular measurement, $r(z)=\int_0^z cH^{-1} dz'$ is the comoving distance to redshift z and $y\equiv \lambda_{21} (1+z)^2/H(z)$ is a conversion factor between frequency intervals and comoving distances ($H(z)$ is the Hubble expansion rate and $\lambda_{21}\approx 21.1$cm the rest frame wavelength of the 21~cm line). We can relate $dk_\perp$ to the resolution in the $u-v$ (visibility) space, $du$ (the Fourier dual of the angular coordinates on the sky) by $dk_\perp=2\pi du/r$ (using our Fourier conventions). Note that
\be 
du=1/\sqrt{A_{FoV}}
\ee
and is related to the effective collecting area of one element of the interferometer, $A_e=\lambda^2 du^2$, where $\lambda$ is the wavelength of observation (but care must be taken in this relation due to multi-beaming). The resolution along the line of sight is just $dk_\parallel=2\pi/({\rm yB})$ (visibilities are Fourier transformed along the frequency coordinate).

Finally, the noise power spectrum is given by
\be
P_N(k,\theta)=r^2 y\frac{T_{sys}^2}{du^2 t(k,\theta)}\,,
\ee
where $t(k,\theta)$ is the time spent observing a given $\bk$ pixel which is just the time spent observing the corresponding pixel of resolution $du^2$ on the $u-v$ plane, $\bu=r \bk_\perp/(2\pi)$. For a total observation time, $t_0$, this will be related to the baseline density distribution, $n(|\bu|)$, through $t(k,\theta)=t_0 du^2 n(|\bu|)=t_0 du^2 n\left(r k sin(\theta)/(2\pi)\right)$ assuming the baseline density is rotationally invariant (which is a good approximation due to Earth rotation). If we make the further assumption that the baseline density distribution is constant on the $u-v$ plane up to a maximum baseline $D_{max}$ (which is not the same as assuming an uniform distribution of antennas), we have
\be
n(|\bu|)=\frac{\lambda^2 N_a^2}{\pi D_{max}^2},
\ee
noting that the integration over half the plane should give approximately $N_a^2/2$ ($N_a$ is the number of elements in the interferometer).
The noise power spectrum then further simplifies to:
\be
P_N(k,\theta)=r^2 y\frac{\pi\lambda^2 D_{max}^2 T_{sys}^2}{t_0 A_{tot}^2}\,,
\ee
where $A_{tot}$ is the total collecting area of the telescope. The system temperature was modeled by
\be
T_{sys}(z)=50+60\left(\frac{1+z}{4.73}\right)^{2.55}\ \ \ {\rm K}
\ee
where the first term is the receiver noise temperature (assumed to be 50 K) and the second term is the much larger sky temperature which is dominated by the Galactic Synchrotron at the redshifts of interest. Note in particular that neither the bandwidth or the frequency resolution show up in the noise calculation (they can be used instead to increase the number of modes $N_m$ available to reduce the total error). Moreover, looking at the last expression we see that $P_N$ decrease as $1/A_{tot}^2$ once we fix $t_0$ and the maximum baseline. Once we achieve $P_S \sim P_N$ the only way to decrease the total error is through $N_m$ which does not depend on the total collecting area or the baseline distribution.

\subsection{Experimental Setup}
\label{setup}

The current generation of radio-interferometers now in construction such as the MWA and LOFAR will not have the collecting area necessary to probe the high redshift universe ($z>15$). This will require a second generation experiment such as the SKA, with an observational window between 70 MHz and 10GHz and planned to be completed by 2020. 
The SKA will probably be made up of three different instruments \citep{schilizzi07, faulkner10}: SKA-low, tailored for the low frequency signal between 70MHz to 450MHz, probably made up of sparse Aperture Arrays so that the collecting area scales as $\lambda^2$; SKA-mid between 400MHz and 1.4 GHz, using dense Aperture Arrays; and SKA-high, using dishes for frequencies above 1.2 GHz (note the small overlap between ranges). In this paper, we will assume a low frequency ``SKA type'' experiment with a setup capable of probing the high redshift, pre-reionization epoch. 

The analysis in the previous Section showed that in order to measure the imprint of the first galaxies on the 21~cm signal, we need to observe at least up to $z=20$, corresponding to a frequency of $\nu\sim 68$MHz. Allowing for some flexibility, we consider an experiment that would be able to measure frequencies down to 60 MHz. Taking into account the design reference for SKA, we assumed a sensitivity of 4000 m$^2$/K at 70 MHz (which in fact requires a total collecting area of $\sim 14\ {\rm Km}^2$, well above the $1 {\rm Km}^2$ used to baptize the telescope) and scaling as $\lambda^2$ around these frequencies. Note that the planned collecting area has been subject to some change along the years and the current design for what is called SKA Phase 1 suggests a sensitivity of 2000 m$^2$/K instead \citep{garrett10}. In order to allow for designs with lower sensitivities, we also consider instruments with 20\% and 10\% of the total collecting area of the design reference SKA.
For the distribution of antennas, the SKADS\footnote{http://www.skads-eu.org} reference design assumes they would be collected in 250 stations of 180 m diameter each, with 66\% in a 5 Km diameter core and the rest along 5 spiral arms out to a 180 Km radius. 

The signal we are looking for dominates on scales $k\lesssim 1$ h/Mpc, requiring baselines up to $\sim 10$ Km for proper sampling at the relevant frequencies, e.g. a resolution of 0.76 arcmin at $z=20$ ($k_{max\perp}\sim 1.8$ h/Mpc - higher resolution can be achieved along the frequency direction although the error would be much larger since we would only have modes along the line of sight). Taking a reasonable setup, we assumed that 70\% of the total collecting area would be concentrated within this core of 10 Km in diameter. We do not use the rest of the collecting area in the noise calculation, assuming instead this would be used for point source removal and calibration. The same applies when the total collecting area is only 10\% and 20\% of the reference design.

For the field of view (FoV) we used 10$\times$10 ${\rm deg}^2$, which, taking $z=20$ as reference, sets a resolution of roughly $dk_\perp\sim 4.5\times 10^{-3}$h/Mpc. Note that the FoV can be traded with the number of beams (in case we want to observe separate fields) and the total instantaneous bandwidth (in case we want to probe a larger redshift range). Assuming that, in the core, correlations within stations are also possible, we considered a minimum baseline of 50 m giving $k_{min\perp} \sim 9\times 10^{-3}$h/Mpc which is enough to probe the scales of interest. The frequency interval for the analysis should be chosen carefully at these frequencies since for a fixed frequency interval, $dz$ increases with redshift. For instance, an interval of 8 MHz used in previous analysis (e.g. \citealt{bowman07}) corresponds to integrating the signal over $dz\sim 2.5$ which will average out important features during the period when \Lyman fluctuations are important. On the other hand, using a small interval means we will have fewer modes along the line of sight. Taking a compromise, we assumed that each measurement was done with an interval of 4 MHz and a resolution of 0.01 MHz, giving $dk_\parallel=0.08$ h/Mpc and $k_{max\parallel}=16.6$ h/Mpc. Note that the total instantaneous bandwidth should be much larger than this ($\sim 50$ MHz), so we can measure several redshifts at the same time. Finally, the total integration time was taken to be 1000 hours (see table~\ref{table_parms} for a summary of the experimental setup).
\begin{table}
\centering                         
\begin{tabular}{c c c}        
\hline\hline                 
Parameters &|& Values \\    
\hline                      
   Min. Freq. &|& 60 MHz \\
   Sensitivity &|& 4000 m$^2$/K \\
   Max. baseline &|& 10 Km \\
   Min. baseline &|& 50 m \\
   FoV &|& 100 deg$^2$ \\
   Integration time &|& 1000 hours \\
   Freq. interval &|& 4 MHz \\
   Freq. resolution &|& 0.01 MHz \\
\hline                                  
\end{tabular}
\caption{Assumed experimental Setup. We also considered 20\% and 10\% of this sensitivity in our analysis. Note that only 70\% of the assumed collecting area is used in the core of 10 Km diameter.}            
\label{table_parms}     
\end{table}

We followed the error calculation described above for the power spectrum assuming a distribution for the antenna/stations of the telescope such that a constant baseline density is obtained. Although this is difficult to achieve in practice since the density of stations normally decreases from the center, it turns out that this configuration gives the best signal to noise in terms of the brightness temperature maps and it should therefore give a good indication of the telescope capabilities to probe the 21~cm signal power spectrum. A final note on foregrounds: they can be the most damaging factor at these low frequencies (together with the calibration issues raised by the ionosphere) although it is expected that if the foregrounds are smooth enough along the frequency direction, they can be simply removed by fitting out a smooth function \citep{jelic08,morales06,santos05}. In this analysis we assume that foreground cleaning was applied on a region of $32$ MHz (so as to avoid edge effects on our $4$ MHz interval) and was successful on scales smaller than this $32$ MHz, e.g. we ignore scales with $k<0.01$ h/Mpc (see \citealt{harker10}). 
Calibration issues raised by the ionosphere can also be quite damaging in particular due to the large field of view of this low frequency interferometers \citep{cohen09, matejek09,liu10} and the situation should become more clear once we have results from the first generation of EoR experiments. In this paper we concentrate on analysing the required minimum setup for an SKA type experiment in order to make a statistical detection of the signal, neglecting for now the challenges posed by the ionosphere.

\subsection{Power Spectrum constraints}

\begin{figure}[!t]
\hspace{-0.5cm}
\includegraphics[scale=0.45]{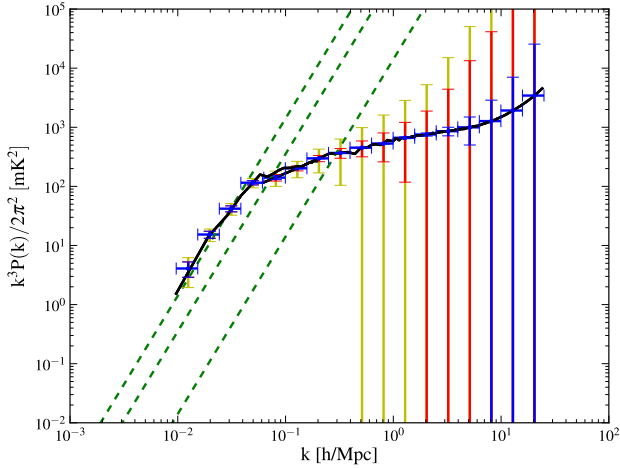}
\caption{Error on the 3d 21cm power spectrum at z=20.25 ($\bar{x}_{\alpha}=0.4$) assuming an SKA type experiment. Solid black - signal for simulations S1+S2; dashed green - noise power spectrum (equation 12) for the full SKA collecting area, 20\% and 10\% (increasing amplitude). The expected error taking into account the available number of modes (equation 7) is shown with error bars in blue, red and yellow respectively. Measurement of the large scale power spectrum should be possible as long as the signal amplitude is not 100 times smaller than current value (this is controlled by $x_\alpha/(1+x_\alpha)(1-T_\gamma/T_S)$ so it should be fine for $x_\alpha > 0.01$). On large scales it should be possible to measure the signal even with only 10\% of the collecting area.}
\label{ps_error}
\end{figure}

Figure \ref{ps_error} shows the expected error on the 3 dimensional power spectrum, $P(k)$ (integrated over $\theta$), assuming the above setup. Even if foreground removal affects larger $k$-modes than assumed here we still have a huge range of measurements available up to $k\sim 7\ $ h/Mpc and in particular, measurements are quite good on the interval where the \Lyman contribution is more important. On very large scales it should be possible to measure the power spectrum even with only 10\% of the collecting area.
Sample variance dominates on large scales while noise (dashed green line) is dominating on small scales (the number of modes on small scales is limited due to the size of the maximum baselines). Note that we could redistribute the collecting area so that the noise power spectrum followed more closely the signal. The error on large scales is kept small because there are still a large number of measurements $N_m$ due to the high resolution on the $u-v$ space from the large field of view (see eq. \ref{ps_error}).
With the above measurements of the 3d power spectra, we should be able to constrain some of the characteristics of the first galaxies such as their emission spectra. Although, as we have seen, the \Lyman signal dominates on larger scales over other contributions, these constraints will be particularly strong if we fix the cosmology, which should be a reasonable assumption given the expected level of constraints from the Planck satellite.

\begin{figure}[!t]
\hspace{-0.5cm}
\includegraphics[scale=0.45]{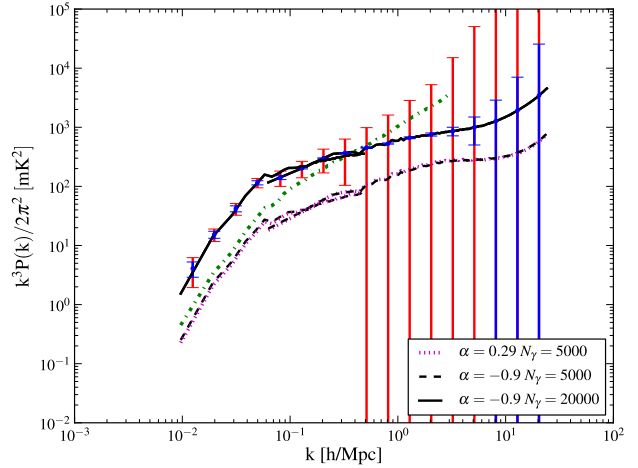}
\caption{21~cm temperature power spectrum for a few Lyman alpha emission models and simulations S1 and S2. Top black solid lines correspond to $\langle x_\alpha \rangle=0.4$ while bottom dashed/dotted curves to $\langle x_\alpha \rangle=0.1$ (all at $z\sim 20.25$). The green dot-dashed line uses halos down to $10^6 \msun$ and a Pop III type emission spectra. The error bars in blue corresponds to the full SKA while the ones in red to 10\% of the collecting area.}
\label{t21_lya}
\end{figure}

Up to now we have been assuming a type II emission spectrum with a total of $20,000$ photons per baryon emitted between the \Lyman and Lyman limit frequency. If instead we assume that the stars responsible for the \Lyman coupling are Pop III, we should have around $5,000$ photons emitted per baryon used in stars with a spectral index of $\alpha=0.29$. Figure \ref{t21_lya} shows the results of changing the emission model with the dashed curves corresponding to the PopIII case (note that due to the lower number of photons, these lines also correspond to a lower $\langle x_\alpha \rangle$ of $0.1$. The dotted curve corresponds to changing the spectral index to $\alpha=0.29$ while maintaining the total emission to $5,000$ photons, which will be difficult to distinguish under the current noise expectations (the total number of emitted photons per baryon is the most relevant parameter for the \lya flux).
Another possibility is to assume that Pop III stars are also being formed in halos with mass less than $10^8 \msun$. The green dot-dashed line in figure \ref{t21_lya} shows the effect of considering halos down to $10^6 \msun$ from simulation S1 at the same redshift. Note that this has a much higher coupling and if we consider higher redshifts so that $\langle x_\alpha \rangle\sim 0.1$ the line will be closer to the dashed one.

These calculations show that it should be possible for SKA to measure \meanxa from observations of the 21~cm power spectrum at $z\approx20$.  Given a detailed model and small error bars this could be measured directly from the shape of the power spectrum.  More generally, if measured over only a small range of wave-numbers, the redshift dependence of \meanxa could be extracted by looking at the redshift evolution of the amplitude and slope in a manner analogous to that suggested by \citet{lidz2008} for reionization. Note that, although during \lya domination different models have similar power spectra for the same \meanxa, its evolution with redshift could be used to distinguish these models.

Measurements of \meanxa in a number of different redshift bins would be instrumental in constructing a ``21 cm Madau plot" \citep{madau1996}.  The Madau plot shows the evolution of the star formation rate as a function of redshift and producing such a plot in the very early Universe would be a major accomplishment of SKA.  Measurements of \meanxa do not give this directly, since in our model \meanxa is determined by the product of the star-formation rate and the emissivity of the galaxies.  Breaking this degeneracy observationally will be difficult and would lead to a plot whose absolute normalisation was uncertain, but whose shape tracked the star-formation rate (assuming no evolution of the galaxies).  However, for many models of early galaxy formation \citep{leitherer1999,bromm2004} the variation in the number of \lya photons produced per baryon is only a factor of two or so.  Thus with a reasonable theoretical prior a useful measure of the star-formation history could be established.

\section{Redshift space distortions}

The signal we are trying to probe is actually anisotropic due to the redshift space distortions set by the peculiar velocity of the HI gas, which is already being taken into account in equation \ref{t21_d}. Using a linear approximation, $A_{\alpha} \approx 
\frac{\langle x_{\alpha}\rangle}{1+\langle x_{\alpha} \rangle}\left(1+\frac{1}{1+\langle x_{\alpha} \rangle}\delta_{x_{\alpha}}\right)$ and $A_{\nu}\approx 1-\frac{1}{H}\frac{dv_r}{dr}$ so to first order the fluctuation in $T_b$ becomes
\be
\delta T_b(\nu) \approx C(z) \left(1+(1+\beta)\delta+\frac{1}{1+\langle x_{\alpha}\rangle}\delta_{x_{\alpha}}-\frac{1}{H}\frac{dv_r}{dr}\right),
\ee

The peculiar velocity field originates from inhomogeneities in the density field, so that in Fourier space we can use Kaiser's approximation $\frac{1}{H}\frac{dv_r}{dr}(\textbf{k})=-\mu^2\delta(\textbf{k})$ \citep{kaiser87} to write the velocity as a function of its dependence in $\mu=\cos(\theta)$. This linear approximation in the velocity contribution allows the brightness temperature power spectra to be separated in only three powers of $\mu$:  
\be
P_{T_b}(k,\mu)= \mu^4 P_{\mu^4}(k)+\mu^2P_{\mu^2}(k)+P_{\mu^0}(k).
\label{Pmu024}
\ee
with
\be
P_{\mu^4}(k)=C^2(z)P_{\delta}(k),
\label{Pmu4}
\ee
\be
P_{\mu^2}(k)=2 C^2(z)\left[(1+\beta)P_{\delta}(k)+\frac{1}{1+\langle x_{\alpha}\rangle}P_{\delta\delta_{x_{\alpha}}}(k)\right],
\label{Pmu2}
\ee
and
\begin{eqnarray}
&&P_{\mu^0}(k)=C^2(z)\Bigg[(1+\beta)^2P_{\delta}(k)+\nonumber\\
&&\left.\left(\frac{1}{1+\langle x_{\alpha}\rangle}\right)^2P_{\delta_{x_{\alpha}}\delta_{x_{\alpha}}}(k)+\frac{2(1+\beta)}{1+\langle x_{\alpha}\rangle}P_{\delta\delta_{x_{\alpha}}}(k)\right],
\label{Pmu0}
\end{eqnarray}
where the power spectra $P_{ab}(k)$ for generic quantities $a$ and $b$ is defined as:
\be
(2\pi)^3\delta^3_D(\bk-\bk')P_{ab}(k)\equiv\frac{\langle a(\bk)b(\bk')^*\rangle+\langle b(\bk)a(\bk')^*\rangle}{2}.
\ee

This decomposition can be particularly useful at the high redshifts we are considering, where, as can be seen from figure \ref{fig_contrib}, the contribution from the velocity compression can be relatively strong. By fitting the observed signal to the above polynomial (equation \ref{Pmu024}) at each $k$, we can in principle separate the cosmological and astrophysical contributions to the brightness temperature fluctuations \citep{barkana05a,barkana05b}. One can use $P_{\mu^2}$ and $P_{\mu^0}$ to learn about the first sources of radiation while the $P_{\mu^4}$ term could be used to measure the matter power spectrum directly, with no interference from any other sources of fluctuations.

\subsection{Assumptions on linearity}

\begin{figure}[!t]  
\includegraphics[scale=0.44]{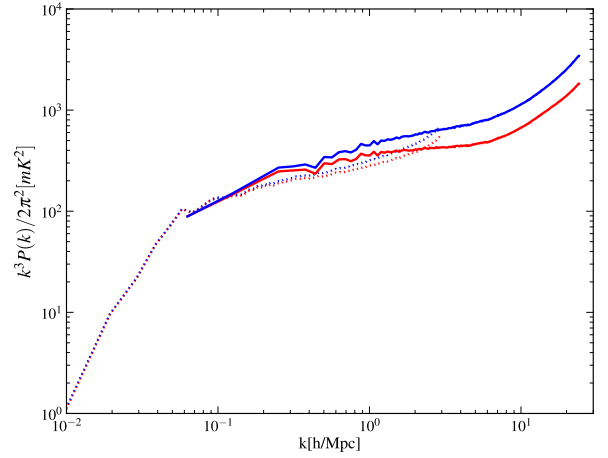}
\caption{Power spectra of brightness temperature fluctuations using simulation S1 (dotted lines) and using simulation S2 (solid lines) for $\langle x_\alpha \rangle=0.4$ ($z\sim 20.25$). Blue lines (top) use the exact velocity contribution and red lines (bottom) use a first order approximation in the velocity contribution.} \label{fig:dtb2_s}
\end{figure}

The decomposition discussed in the previous section relies on assumptions of linearity in both the velocity field and the \lya fluctuations.  On small scales, both of these assumptions may break down and we consider both in turn.
As we go to small scales, fluctuations on the density field increase and the
non-linearities in the velocity contribution can become important, so that the first order approximation to the full velocity term in equation \ref{t21} breaks down. Figure \ref{fig:dtb2_s} compares the brightness temperature power spectrum from the full calculation with the case where we assume the first order approximation for the velocity.
We see that for $k\gtrsim 1.0$ h/Mpc the non-linearities can become important.
This result implies that for smaller scales the correct brightness temperature power spectra has a much more complicated dependence in $\mu$ than the one given by equation \ref{Pmu024} which will complicate the model independent extraction of the astrophysical contributions \citep{shaw2008}.

\begin{figure}[!t]
\hspace{-0.5cm}
\includegraphics[scale=0.45]{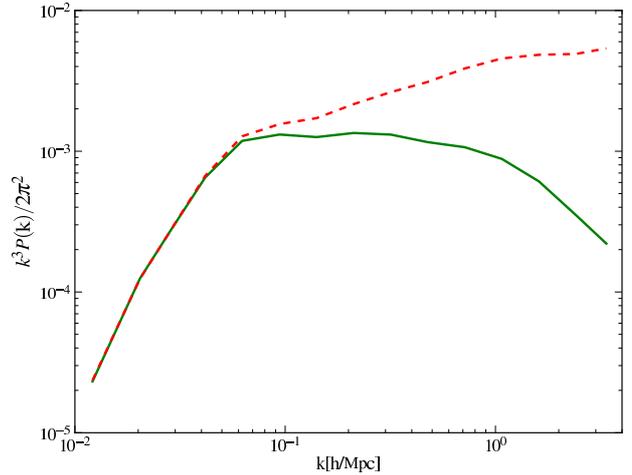}
\caption{Power spectra of fluctuations in $x_{\alpha}/(1+x_{\alpha})$ built using simulation S1 for $\langle x_\alpha \rangle=0.4$ ($z\sim 20.25$). The figure shows the the total fluctuation (solid green line) and its first order approximation (dashed red line).}
\label{fig:ps_xa_2o2}
\end{figure}

Fluctuations in the \lya background too can leave the linear regime.  The information from the Ly$\alpha$ background and consequently from the physical processes that originated it, is contained in ${\delta_{x_{\alpha}}}$ and it can only be directly obtained from the power spectrum angular separation in powers of $\mu$ for scales where the linear approximation of $A_{\alpha}$ is valid.
To show the limits where the first order approximation can be used, we plotted in figure \ref{fig:ps_xa_2o2} the power spectra of $A_{\alpha}$ compared to its linear expression, using the values from simulation S1. The linear approximation should be valid for $k\lesssim 0.2$ h/Mpc.

\begin{figure}[!t] 
\hspace{-0.5cm} 
\includegraphics[scale=0.44]{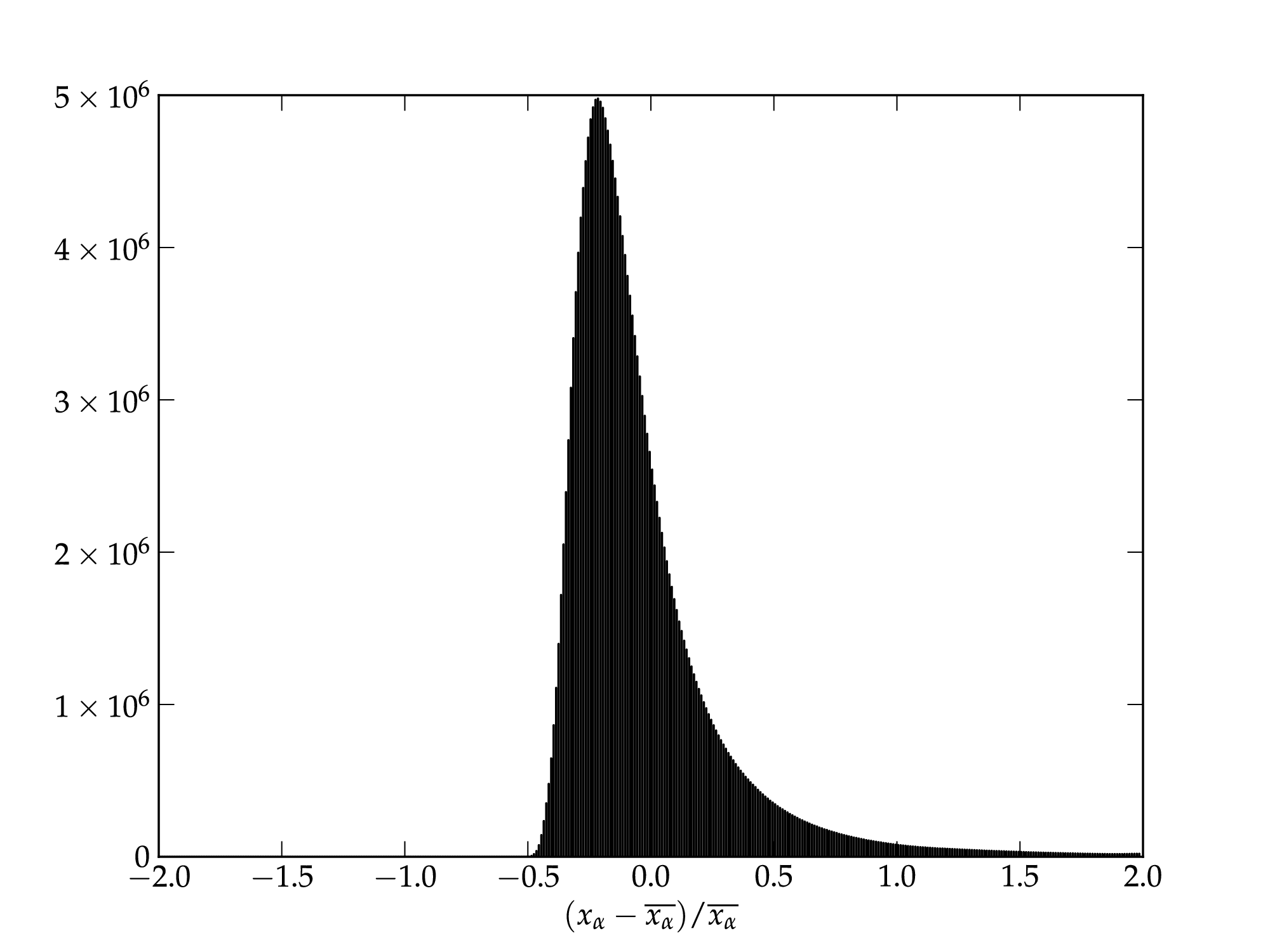}
\caption{Detail of the histogram of fluctuations in $\delta_{x_{\alpha}}$ using simulation S1 for $\langle x_\alpha \rangle=0.4$ ($z\sim 20.25$). The fluctuations in $\delta_{x_{\alpha}}$ go from -50 till 350} \label{fig:hist_xaxm}
\end{figure}

Looking at the histogram of $\delta_{x_\alpha}$ in figure \ref{fig:hist_xaxm} we see that there are few points where $\delta_{x_{\alpha}}>1$. Although this number is small compared to the overall distribution (most cells in the simulation have actually $x_\alpha\sim 0$ at these high redshifts), they are enough to introduce non-linearities on the power spectrum calculation.
This basically implies that for small scales the fluctuations in $x_{\alpha}$ will have a large dependence on the random distribution of the first galaxies so it will only be possible to do a linear approximation on $A_{\alpha}$ for scales large enough that a few small peaks of large intensity do not dominate the power spectra of the fluctuation.
Some of the strong coupling regions coincide with ionized regions which may diminish the effect of strong fluctuations in the brightness temperature power spectra, however according with simulations S1, S2 and S3, at the high redshifts we are interested ($z>20$) the ionized regions are so small that including this quantity has no meaningful effect in the linear approximation used.\\

From this study we can conclude that the brightness temperature power spectrum angular separation in equation \ref{Pmu024} should be reasonable for $k< 1$ h/Mpc and the linear approximations in equations \ref{Pmu2} and \ref{Pmu0} should be valid for larger scales, $k\lesssim 0.2$ h/Mpc, allowing the direct extraction of information about the \Lyman field \citep{barkana05b}. This is fortunate since it corresponds to the scales where the \Lyman fluctuations dominate the contributions to the brightness temperature (figure \ref{fig_contrib}) as well as the scales where the SKA should perform better (figure \ref{ps_error}). For $0.2 \lesssim k \lesssim 1.0$ h/Mpc we can still use the angular decomposition but equations \ref{Pmu2} and \ref{Pmu0} will include higher order terms. Given the scales we are interested in, from now on we will concentrate on simulation S1 ($1000$ Mpc) which can be used to calculate the power spectrum up to $k\sim 2.3$ h/Mpc.

\subsection{Constraints on P$(k,\mu)$}

\begin{figure}[!t]  
\includegraphics[scale=0.47]{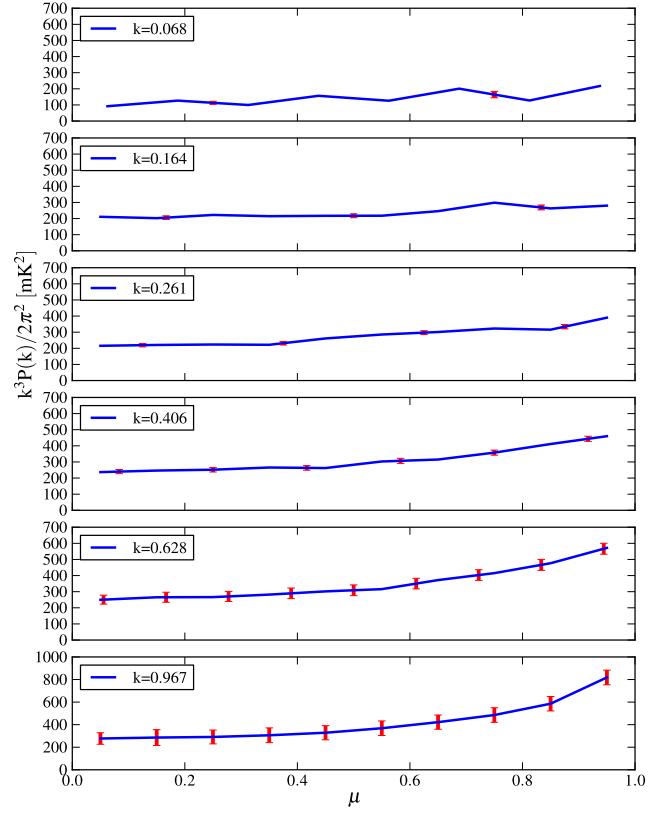}
\vspace{-1.0cm} 
\caption{Expected measurements for the SKA of the 21~cm power spectrum as a function of the angle with the line of sight ($z=20.25$, $\bar{x}_\alpha=0.4$).}
\label{ps_mu}
\end{figure}

As we have seen in the previous section, the redshift space distortions introduce an anisotropy in the 21~cm power spectrum that, in the linear regime, depends on $k$ and even powers of $\mu$. More generally non-linearities in the velocity field will introduce terms with more complicated dependence \citep{shaw2008}. 
Measurements of P$(k,\mu)$ might provide more information on the \lya signal \citep{barkana05a} than the spherical average P$(k)$ we considered so far.
Figure \ref{ps_mu} shows the expected errors in the measurements of the 3d power spectrum as a function of $\mu$, for an SKA type experiment (table~\ref{table_parms}).
The calculation follows equation~\ref{pk_error} where the number of modes in each bin is now a function of $k$ and $\mu$. We used bins with a logarithmic spacing in $k$ and linear in $\mu$ (with $0\leq \mu \leq 1$ since the power spectrum only depends on $\mu^2$). The number of modes in each bin is accounted for by gridding the $\bk$ space with a resolution along the line of sight set by the interval in frequency used for the analysis and a resolution perpendicular to the line of sight set by the field of view (see section \ref{setup}). The minimum size of the bins is given by this resolution. Note that the number of independent $\mu$ bins increase with $k$, although, as can be seen from figure~\ref{ps_mu}, the errors will also increase due to the experimental noise. Moreover, the power spectrum is quite flat for small $k$ which will make it harder to extract the polynomial dependence (this is because the $P_{\mu^0}$ term is dominating in equation~\ref{Pmu024}).

\begin{figure}[!t]
\hspace{-0.5cm}
\includegraphics[scale=0.45]{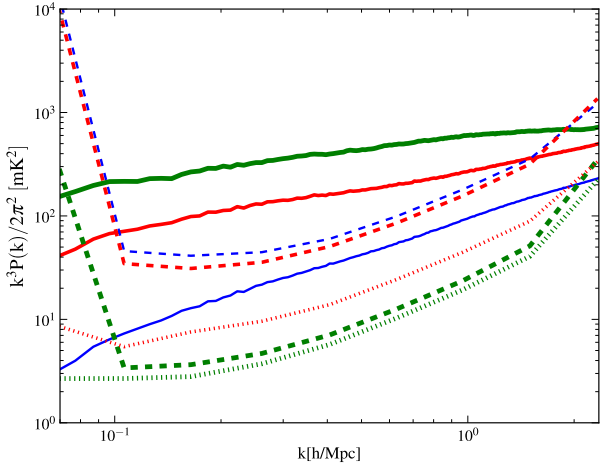}
\caption{Contributions for the 21~cm temperature power spectrum using simulations (solid lines), built using simulation S1 for $\langle x_\alpha \rangle=0.4$ ($z\sim 20.25$). Dashed lines: expected errors. Dotted lines: expected errors assuming that $P_{\mu^4}$ is known. Solid lines from top to bottom are $P_{\mu^0}$ (green), $P_{\mu^2}$ (red) and $P_{\mu^4}$ (blue).}
\label{fish_pmu}
\end{figure}

Although difficult as we have seen, we can try to obtain $P_{\mu^0}$, $P_{\mu^2}$ and $P_{\mu^4}$ by fitting equation \ref{Pmu024} to the measured $P(k,\mu)$ for $k\lesssim 1.0$ h/Mpc. This in turn should allow us to obtain direct constraints on a combination of $P_{\delta_{x_\alpha}\delta_{x_\alpha}}$ and $P_{\delta\delta_{x_\alpha}}$. For small $k$ there are fewer bins so the uncertainty in $P_{\mu^n}$ will be high even for those scales where the errors in $P(k,\mu)$ are small. 
The expected errors on the $P_{\mu^n}$ term can be obtained using a Fisher matrix approach \citep{fisher35}. For parameters $a,b=\{P_{\mu^0}, P_{\mu^2}, P_{\mu^4}\}$, the matrix is calculated through
\be
F_{ab}(k_i)=\sum_j{\frac{1}{\Delta P(k_i,\mu_j)^2}\frac{\partial P(k_i,\mu_j)}{\partial a}\frac{\partial P(k_i,\mu_j)}{\partial b}}
\ee
where again $P(k,\mu)$ is given by equation~\ref{Pmu024} and $\Delta P(k_i,\mu_j)$ is the error for each bin. The covariance of the $P_{\mu^n}$ terms is just the inverse of the Fisher matrix and figure~\ref{fish_pmu} shows the expected 1-sigma errors ($\sqrt{(F^{-1})_{aa}}$).
We see that it should be possible to measure with reasonable accuracy both the $P_{\mu^0}$ and $P_{\mu^2}$ terms which can already give us interesting constraints on the \Lyman field. In fact, if we fix the cosmology (e.g. $P_\delta$), then we should be able to extract information from $P_{\delta\delta_{x_\alpha}}$ and $P_{\delta_{x_\alpha}\delta_{x_\alpha}}$ separately.
On the other hand, the $P_{\mu^4}$ will be hard to measure, which will make the extraction of cosmological information more difficult, simply because it is an order of magnitude smaller than the other terms on the scales we are focusing here. 

In order to improve this last constraint we would have to increase the collecting area of the instrument and the frequency interval used ($4$ MHz in this case) so to have more modes along the line of sight (note however that this will lead to cosmological evolution of the signal along the frequency bin, as already discussed). The SKA should be able to measure modes up to $k\sim 10$ h/Mpc which could have higher values of $P_{\mu^4}$, however for $k>1$ h/Mpc the angular decomposition is no longer valid and the separate measurements of the $P(k,\mu)$ terms will be of little use (except maybe for foreground removal).

In this case, it might be better to just try to fit the parameters of the model directly to the averaged power spectrum $P(k)$ using the simulations (as shown in figure \ref{t21_lya}). If on the other hand we assume we already know $P_{\mu^4}$ with reasonable accuracy then the errors on $P_{\mu^2}$ will improve considerably (figure \ref{fish_pmu}) and any degeneracies between $P_{\mu^0}$ and $P_{\mu^2}$ will be broken.

\subsection{Extracting the Poisson term}

Finally, by measuring the $P(k,\mu)$ terms, we can in principle use an estimator to probe the properties of the first sources of radiation directly through the \Lyman field, without having to go through a full, model dependent, parameter fit to the data. This is based on the decomposition proposed in \citet{barkana05a}, valid in the linear regime. In that case, $\delta_{x_{\alpha}}$ can be expressed as a function of its dependence on effects correlated and uncorrelated with $\delta$ as:
\be
\delta_{x_{\alpha}}(\bk)=W(k)\delta(\bk)+\delta_p(\bk).
\label{x_w_p}
\ee
Note that this decomposition is actually quite general and the main assumption here is the linearity on $\delta$.
$\delta_P$ is the Poisson contribution that arises from the statistical fluctuations in the number density of the rare first galaxies (galaxies being discrete objects and therefore not tracking the continuous density field perfectly) and which to first order is uncorrelated with the density fluctuations. 

The window function, W(k), contains several effects by which $\delta_{x_{\alpha}}$ is related to the underlying matter density fluctuations, $\delta$ (see \citealt{barkana05b} for details). The strongest effect between $\delta$ and $\delta_{x_{\alpha}}$ is the biasing of the number density of galaxies  with respect to the density fluctuations by a factor $b(z)$ (the average halo bias) so that an overdense region will have a factor ($1+b(z)\delta$) more sources \citep{mo96}. This is the only term that is implicitly included in our simulations through the way we do the 3d integration of the \lya flux. Note however that this term completely dominates over the other terms in \citet{barkana05b} since $b(z) \sim 15$ at $z\sim 20$.
In that case, we can express W(k) as:
\be\label{W_t}
W(k)=\frac{1}{\langle x_{\alpha} \rangle}\int^{z_{max(2)}}_z dz' \frac{D(z')}{D(z)}\frac{dx_{\alpha}}{dz'}b(z') \frac{\sin{(kr)}}{kr},
\ee
where $D(z)$ is the linear growth function and
we are just considering photons emitted below the Lyman beta limit for comparison, so that
$z_{max(2)}=(32/27)(1+z)-1$ \citep{barkana05b,pritchard06}.

\begin{figure}[!t]  
\hspace{-0.5cm} 
\includegraphics[scale=0.44]{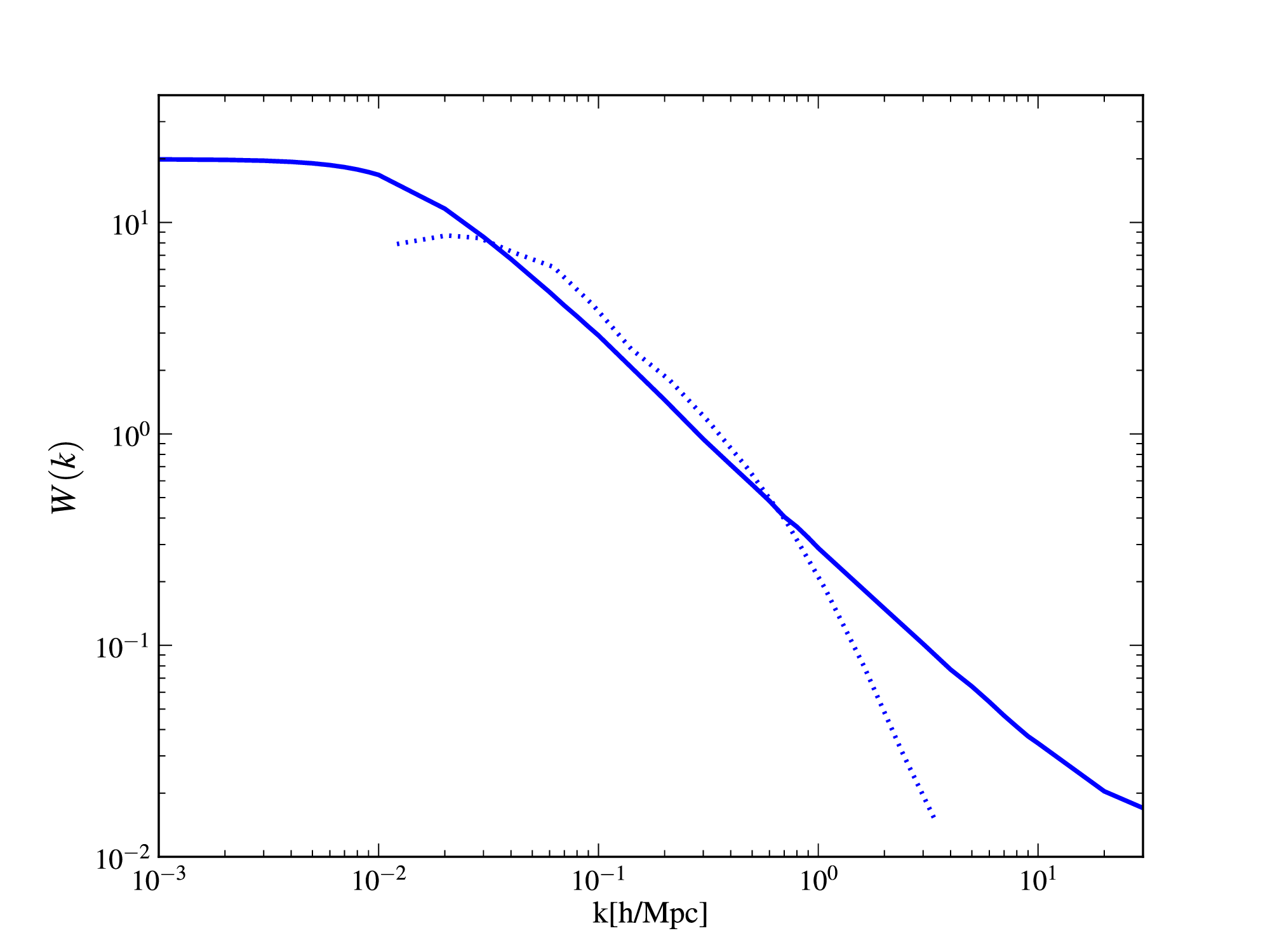}
\caption{Window function for simulation S1 considering only photons emitted below the Lyman beta limit, for $\langle x_\alpha \rangle=0.4$ ($z\sim 20.25$). Solid lines are obtained using equation \ref{W_t} and dotted lines are obtained using $P_{\delta\delta_{x_{\alpha}}}(k)$ and $P_{\delta}(k)$.} 
\label{fig:Wt}
\end{figure}
Using the parametrization of $\delta_{x_{\alpha}}$ we can write $P_{x_{\alpha}}(k)=W^2(k)P_{\delta}(k)+ P_p(k)$ and $P_{\delta\delta_{x_{\alpha}}}(k)=W(k)P_{\delta}(k)$, meaning that we can isolate $W(k)$ from the simulations ($W(k)=P_{\delta\delta_{x_{\alpha}}}(k)/P_{\delta}(k)$). 
In figure \ref{fig:Wt} we show the window function obtained from equation \ref{W_t}, with values from simulation S1, which has the same behavior as the one computed by \citet{pritchard06}. Figure \ref{fig:Wt} also shows that the window function obtained from equation \ref{W_t} has a higher amplitude for large scales and small scales than the one obtained from simulation S1 using $P_{\delta}$ and $P_{\delta\delta_{x_{\alpha}}}$. On large scales this is probably a consequence of our box still not being quite large enough to encompass the rarest, most highly biased objects leading to an underestimate of the power on large scales.  On small scales, this seems to be a consequence of the onset of non-linearities in the density and \lya fields.  On scales close enough to clusters of sources that the \lya coupling saturates, we might expect to see a steeper fall off in the 21 cm power as seen here. Nevertheless, $W(k)$ obtained from the simulation follows quite closely the expected theoretical values, which supports the decomposition given in equation~\ref{x_w_p} and validates the signal extraction we discuss next.

\begin{figure}[!t]  
\hspace{-0.5cm} 
\includegraphics[scale=0.44]{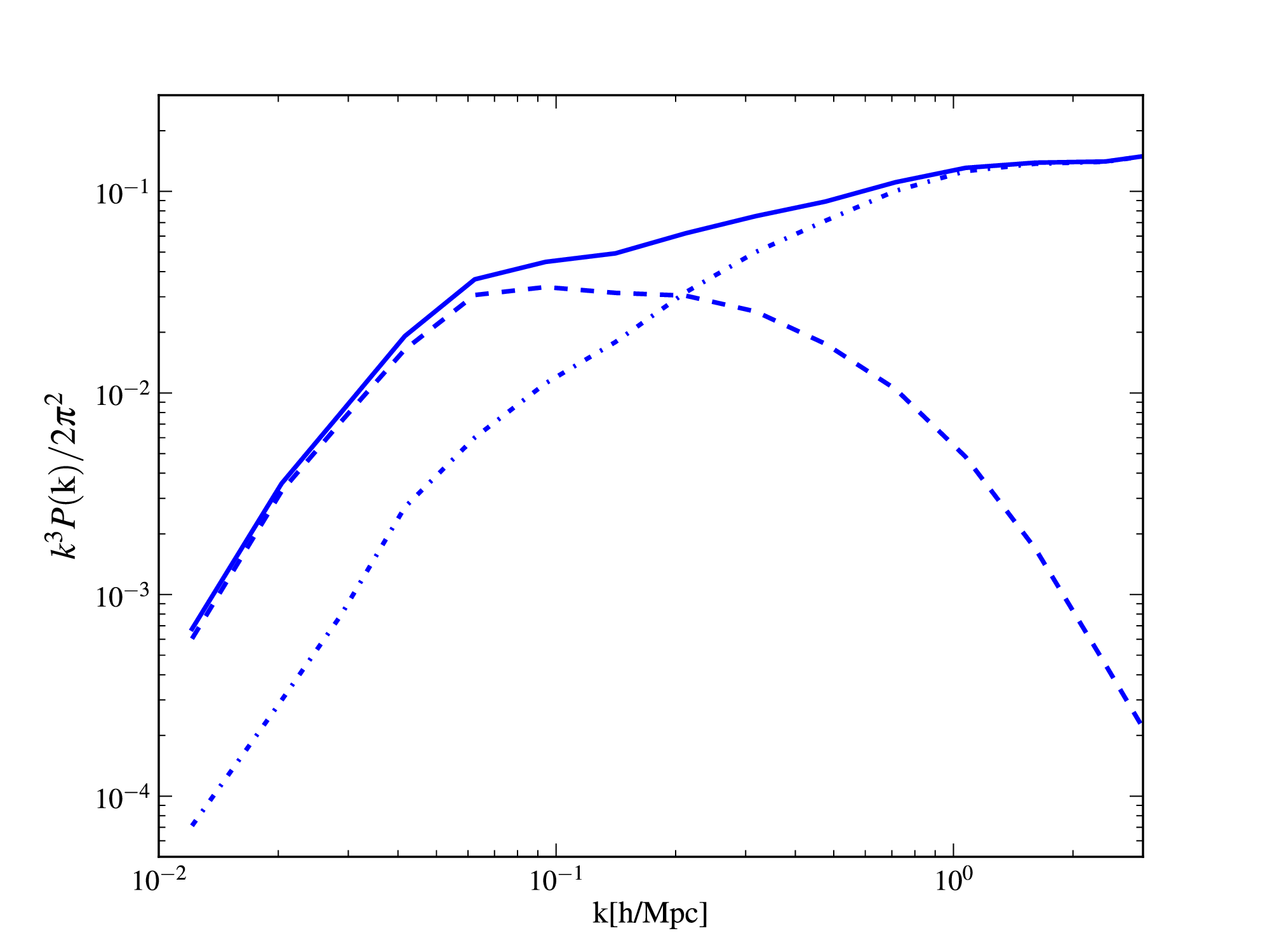}
\caption{Power spectra of fluctuations in $x_{\alpha}$ using simulation S1 for $\langle x_\alpha \rangle=0.4$ ($z\sim 20.25$). Shown are $P_{x_{\alpha}}$ (solid line), $W^2P_{\delta}$ (dashed line) and $P_p$(dotted line).} 
\label{fig:poisson_xa}
\end{figure}
Also, in figure \ref{fig:poisson_xa} we show $P_{\delta_{x_{\alpha}}}$, $W^2P_{\delta}$ and $P_p$, built with $P_{\delta}$, $P_{\delta_{x_{\alpha}}}$ and $P_{\delta\delta_{x_{\alpha}}}$ from simulation S1. This shows the behavior expected from the analytic model of \citet{barkana05b} and gives an amplitude of $P_p$ that is broadly consistent with the values that we calculate from their model.  The Poisson contribution dominates for $k>0.2$ h/Mpc so that there is just a small interval where we can expect to probe this term before non-linear effects correlated to the density field become relevant. Note that the Poisson contribution flattens on small scales $k\gtrsim1 {\rm(h/Mpc)}$ slightly more than predicted by the analytic model.  This is similar to the drop off in power on small scales seen in $W(k)$ and probably occurs for similar reasons.

For large enough scales, we can use the observational data to obtain the window function through
\be
\frac{P_{\mu^2}(k)}{P_{\mu^4}(k)}=2\times\left[1+\beta+\frac{1}{1+\langle x_{\alpha}\rangle}W(k)\right],
\label{W_obs}
\ee
and the Poisson contribution uncorrelated with $\delta$:
\be
P_{un-\delta}(k)=P_{\mu^0}(k)-\frac{P^2_{\mu^2}(k)}{4P_{\mu^4}(k)}=C^2(z)\left(\frac{1}{1+\langle x_{\alpha} \rangle}\right)^2P_p(k).
\label{P_p}
\ee

\begin{figure}[!t]  
\hspace{-0.5cm} 
\includegraphics[scale=0.44]{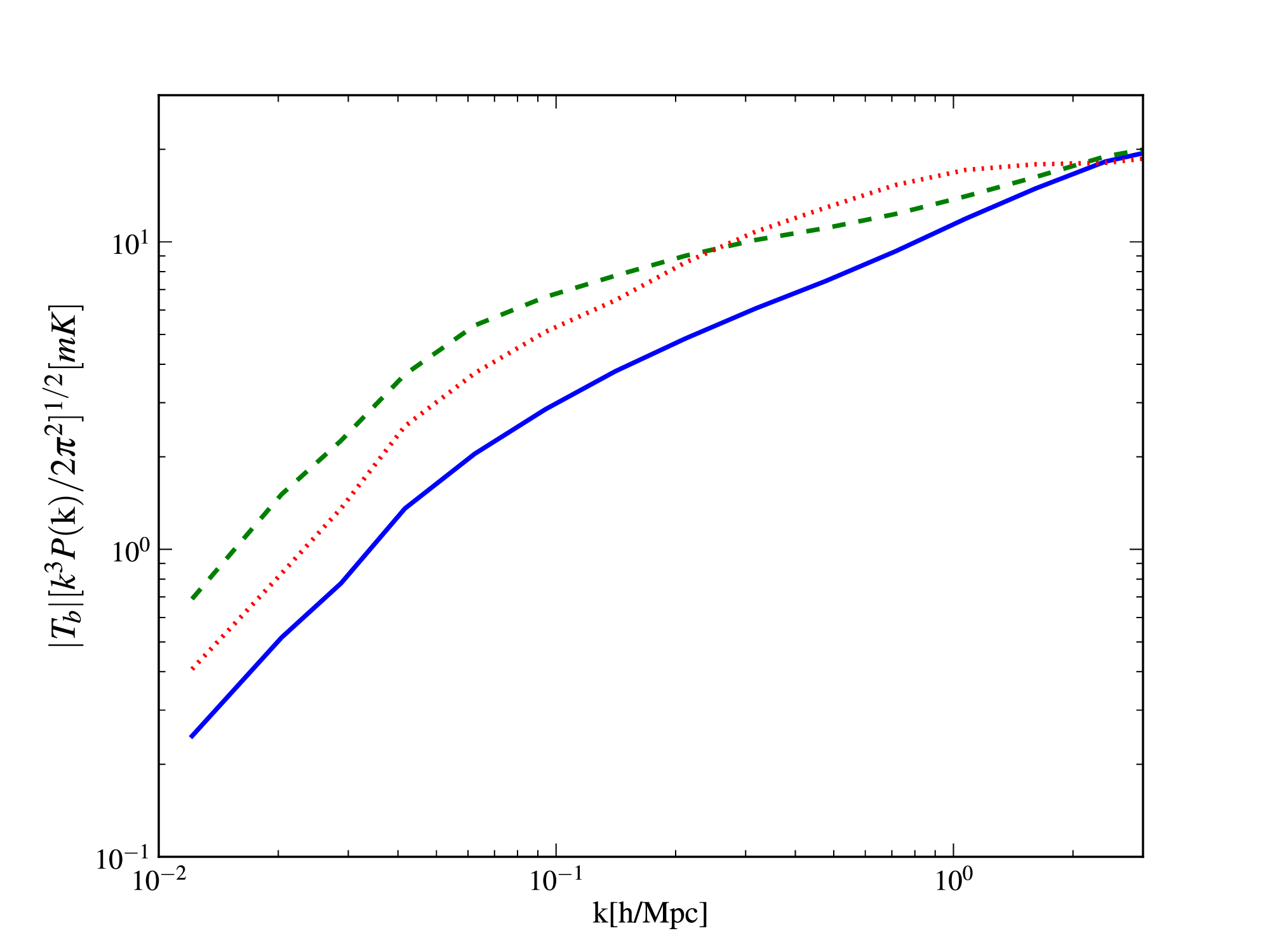}
\caption{Power spectra of brightness temperature fluctuations using simulation S1 for $\langle x_\alpha \rangle=0.4$ ($z\sim 20.25$). Shown are $P_{\mu^2}$ (green dashed line) $2 C(z) P_{\delta}$ (blue solid line), and $P_{un-\delta}$ (red dotted line).} 
\label{fig:pmu_mk}
\end{figure}
The use of equations \ref{W_obs} and \ref{P_p} allows us to characterize the \lya background and the number and distribution of galaxies at these early redshifts.
In figure \ref{fig:pmu_mk}, we used simulation S1 to show $P_{un-\delta}$ and $P_{\mu^2}$, which are consistent with the results obtained by \citet{barkana05b}.
\begin{figure}[!t]
\hspace{-0.5cm}
\includegraphics[scale=0.45]{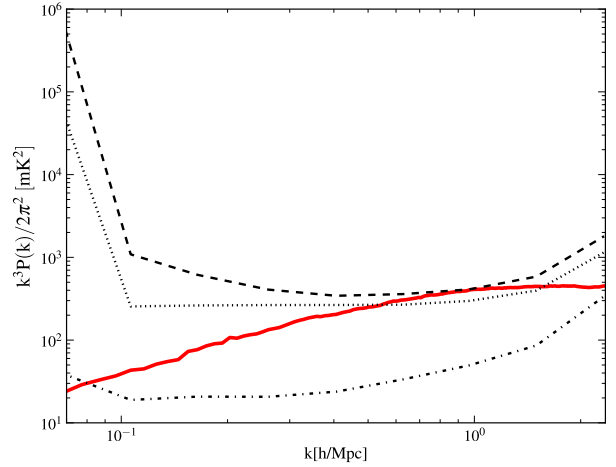}
\caption{Power spectra of $C^2(z)\left(1/(1+\langle x_{\alpha} \rangle)\right)^2 P_p(k)$ (solid line), built using simulation S1 for $\langle x_\alpha \rangle=0.4$ ($z\sim 20.25$). Expected error (dashed line) and expected error assuming that $P_{\delta}$ is known (dotted line), so the error in $P_{\mu^4}$ comes only from $C(z)$. Dot-dashed line assumes that $C(z)$ is known to high accuracy.}
\label{fish_corr2}
\end{figure}
Figure \ref{fish_corr2} shows the power spectra of the Poisson contribution (equation \ref{P_p}) and the expected errors associated with this measurement. As expected, due to the high error in $P_{\mu^4}$, the measurement of this Poisson term will be quite difficult even with the assumed experimental setup (dashed line). If we assume that the matter power spectrum is known with reasonable accuracy from CMB data and galaxy surveys, then we can combine the measurement of the $P_{\mu^4}$ term at several $k$ to constrain $C^2(z)$ (eq. \ref{Pmu4}), e.g. the mean brightness temperature on the sky. Unfortunately the constraint is still large, $C^2(z)\approx (6.42\pm 5.60)\times 10^3 {\rm mK}^2$, which translates into large errors for the Poisson term (dotted lines). Nevertheless, a detection will still be possible, with a signal to noise of 2.6 (dashed line) and 3.8 (dotted line).

If we further assume that we combine this result with the measurements from a global experiment in order to measure the mean brightness temperature with high accuracy, then we can extract all the relevant power spectra reasonably well (dot-dashed line).

\section{Conclusions}

In this paper, we have made use of a recently developed fast semi-numerical code to explore the possibility for 21 cm observations of the time of the first galaxies with SKA.  We demonstrated that this code allows the simulation of the full range of scales likely to be accessible to observations.  Here we have focused on the impact of 21 cm fluctuations from large fluctuations in the \lya coupling, which arise due to the clustering of the rare first galaxies and considered different emission models.

Measurement of these fluctuations would provide insight into the formation of the very first galaxies and would be highly complementary with the next generation of large optical/IR telescopes.  We have shown that SKA-pathfinders with $\sim10\%$ of the full collecting area should be capable of making a statistical detection of the 21 cm power spectrum at redshifts $z\lesssim20$.  With the full SKA sensitivity this detection would become a measurement allowing astrophysical properties of the first galaxies to be determined. Note that it is crucial that these experiments are able to go below 70 MHz in order to probe the signal.

Observations with an SKA-like instrument would enable the determination of \meanxa as a function of redshift from the amplitude and shape of the 21~cm fluctuations. Although different models with the same \meanxa give similar power spectra, this evolution with redshift should allow to distinguish them. Even though there is a strong degeneracy between the star formation rate and the UV spectral properties of the galaxies themselves, these observations would enable constraints to  be placed on the star formation rate in the earliest generations of galaxies.  Although crude, the resulting ``Madau plot" would be very useful for understanding the formation processes of the first galaxies.

Moving beyond the angle-averaged power spectrum, we have investigated the use of redshift-space distortions to separate out different components of the power spectrum as suggested by \citet{barkana05b}.  These measurements are difficult due to instrumental limitations and are further complicated by non-linearities on small scales.  For the first time, we have investigated the non-linearity of \lya fluctuations showing that they should be important on scales $k\gtrsim 0.2{\rm(h/Mpc)}$.  Nonetheless, on larger scales the detection of these effects is possible with SKA level sensitivity and should allow to put direct constraints on the \lya power spectrum.

Redshift space distortions further offer the possibility of extracting components of the power spectrum that do not correlate with the density field.  This Poissonian component contains direct information about the number density of sources and so gives complementary information to the other sources of fluctuations. We have shown that a detection is possible and if strong assumptions about the underlying cosmology are possible and combined with information about the mean 21 cm signal, then these fluctuations may be picked out clearly by SKA.

These results illustrate the potential of 21 cm observations to shed new light on the astrophysics during the pre-Reionization epoch.  In the future, our improved knowledge of cosmological parameters will provide a firm foundation to pick out the details of galaxy formation in the early Universe. SKA and pathfinders capable of observing at frequencies $\nu\lesssim100{\rm\,MHz}$ will begin to access this interesting period and transform our understanding of the cosmic dawn.

\begin{acknowledgements}

This work was partially supported by FCT-Portugal under grants PTDC/FIS/66825/2006 and PTDC/FIS/100170/2008.
JRP is supported by NASA through Hubble Fellowship grant HST-HF-01211.01-A
awarded by the Space Telescope Science Institute, which is operated by the
Association of Universities for Research in Astronomy, Inc., for NASA,
under contract NAS 5-26555. AC acknowledges support of NSF grant CAREER AST-0645427. MGS was a visitor at UCI
when this work was concluded.

\end{acknowledgements}

\bibliographystyle{aa}
\bibliography{Lya21cm}

\begin{thebibliography}{63}
\expandafter\ifx\csname natexlab\endcsname\relax\def\natexlab#1{#1}\fi

\bibitem[{{Baek} {et~al.}(2009){Baek}, {Di Matteo}, {Semelin}, {Combes}, \&
  {Revaz}}]{baek09}
{Baek}, S., {Di Matteo}, P., {Semelin}, B., {Combes}, F., \& {Revaz}, Y. 2009,
  \aap, 495, 389

\bibitem[{{Baek} {et~al.}(2010){Baek}, {Semelin}, {Di Matteo}, {Revaz}, \&
  {Combes}}]{baek10}
{Baek}, S., {Semelin}, B., {Di Matteo}, P., {Revaz}, Y., \& {Combes}, F. 2010,
  ArXiv e-prints

\bibitem[{{Barkana} \& {Loeb}(2001)}]{barkana2001}
{Barkana}, R. \& {Loeb}, A. 2001, \physrep, 349, 125

\bibitem[{{Barkana} \& {Loeb}(2004)}]{2004BarkanaLoeb}
{Barkana}, R. \& {Loeb}, A. 2004, \apj, 609, 474

\bibitem[{{Barkana} \& {Loeb}(2005{\natexlab{a}})}]{barkana05a}
{Barkana}, R. \& {Loeb}, A. 2005{\natexlab{a}}, \apjl, 624, L65

\bibitem[{{Barkana} \& {Loeb}(2005{\natexlab{b}})}]{barkana05b}
{Barkana}, R. \& {Loeb}, A. 2005{\natexlab{b}}, \apj, 626, 1

\bibitem[{{Bowman} {et~al.}(2006){Bowman}, {Morales}, \& {Hewitt}}]{bowman2006}
{Bowman}, J.~D., {Morales}, M.~F., \& {Hewitt}, J.~N. 2006, \apj, 638, 20

\bibitem[{{Bowman} {et~al.}(2007){Bowman}, {Morales}, \& {Hewitt}}]{bowman07}
{Bowman}, J.~D., {Morales}, M.~F., \& {Hewitt}, J.~N. 2007, \apj, 661, 1

\bibitem[{{Bromm} \& {Larson}(2004)}]{bromm2004}
{Bromm}, V. \& {Larson}, R.~B. 2004, \araa, 42, 79

\bibitem[{{Chuzhoy} \& {Zheng}(2007)}]{chuzhoy2007}
{Chuzhoy}, L. \& {Zheng}, Z. 2007, \apj, 670, 912

\bibitem[{{Cohen} \& {R{\"o}ttgering}(2009)}]{cohen09}
{Cohen}, A.~S. \& {R{\"o}ttgering}, H.~J.~A. 2009, \aj, 138, 439

\bibitem[{{Cohn} \& {White}(2008)}]{2008CohnWhite}
{Cohn}, J.~D. \& {White}, M. 2008, \mnras, 385, 2025

\bibitem[{{Cuby} {et~al.}(2007){Cuby}, {Hibon}, {Lidman}, {Le F{\`e}vre},
  {Gilmozzi}, {Moorwood}, \& {van der Werf}}]{2007Cuby}
{Cuby}, J., {Hibon}, P., {Lidman}, C., {et~al.} 2007, \aap, 461, 911

\bibitem[{{Faulkner}(2010)}]{faulkner10}
{Faulkner}, A. J. e.~a. 2010, SKADS Memos: http://www.skads-eu.org

\bibitem[{{Field}(1959)}]{field1959spint}
{Field}, G.~B. 1959, \apj, 129, 536

\bibitem[{{Fisher}(1935)}]{fisher35}
{Fisher}, R. 1935, J.Roy.Stat.Soc., 98, 39

\bibitem[{{Furlanetto} {et~al.}(2006){Furlanetto}, {Oh}, \&
  {Briggs}}]{furlanetto06c}
{Furlanetto}, S.~R., {Oh}, S.~P., \& {Briggs}, F.~H. 2006, \physrep, 433, 181

\bibitem[{{Garrett} {et~al.}(2010){Garrett}, {Cordes}, {Deboer}, {Jonas},
  {Rawlings}, \& {Schilizzi}}]{garrett10}
{Garrett}, M.~A., {Cordes}, J.~M., {Deboer}, D.~R., {et~al.} 2010, ArXiv
  e-prints

\bibitem[{{Harker} {et~al.}(2010){Harker}, {Zaroubi}, {Bernardi}, {Brentjens},
  {de Bruyn}, {Ciardi}, {Jeli{\'c}}, {Koopmans}, {Labropoulos}, {Mellema},
  {Offringa}, {Pandey}, {Pawlik}, {Schaye}, {Thomas}, \&
  {Yatawatta}}]{harker10}
{Harker}, G., {Zaroubi}, S., {Bernardi}, G., {et~al.} 2010, \mnras, 405, 2492

\bibitem[{{Heitmann} {et~al.}(2007){Heitmann}, {Lukic}, {Fasel}, {Habib},
  {Warren}, {White}, {Ahrens}, {Ankeny}, {Armstrong}, {O'Shea}, {Ricker},
  {Springel}, {Stadel}, \& {Trac}}]{2007Heitmann}
{Heitmann}, K., {Lukic}, Z., {Fasel}, P., {et~al.} 2007, ArXiv e-prints, 706

\bibitem[{{Hibon} {et~al.}(2009){Hibon}, {Cuby}, {Willis}, {Cl{\'e}ment},
  {Lidman}, {Arnouts}, {Kneib}, {Willott}, {Marmo}, \& {McCracken}}]{2009Hibon}
{Hibon}, P., {Cuby}, J., {Willis}, J., {et~al.} 2009, ArXiv e-prints

\bibitem[{{Hirata}(2006)}]{hirata2006lya}
{Hirata}, C.~M. 2006, \mnras, 367, 259

\bibitem[{{Jeli{\'c}} {et~al.}(2008){Jeli{\'c}}, {Zaroubi}, {Labropoulos},
  {Thomas}, {Bernardi}, {Brentjens}, {de Bruyn}, {Ciardi}, {Harker},
  {Koopmans}, {Pandey}, {Schaye}, \& {Yatawatta}}]{jelic08}
{Jeli{\'c}}, V., {Zaroubi}, S., {Labropoulos}, P., {et~al.} 2008, \mnras, 389,
  1319

\bibitem[{{Kaiser}(1987)}]{kaiser87}
{Kaiser}, N. 1987, \mnras, 227, 1

\bibitem[{{Komatsu} {et~al.}(2010){Komatsu}, {Smith}, {Dunkley}, {Bennett},
  {Gold}, {Hinshaw}, {Jarosik}, {Larson}, {Nolta}, {Page}, {Spergel},
  {Halpern}, {Hill}, {Kogut}, {Limon}, {Meyer}, {Odegard}, {Tucker}, {Weiland},
  {Wollack}, \& {Wright}}]{komatsu2010}
{Komatsu}, E., {Smith}, K.~M., {Dunkley}, J., {et~al.} 2010, ArXiv e-prints

\bibitem[{{Leitherer} {et~al.}(1999){Leitherer}, {Schaerer}, {Goldader},
  {Gonz{\'a}lez Delgado}, {Robert}, {Kune}, {de Mello}, {Devost}, \&
  {Heckman}}]{leitherer1999}
{Leitherer}, C., {Schaerer}, D., {Goldader}, J.~D., {et~al.} 1999, \apjs, 123,
  3

\bibitem[{{Lewis} {et~al.}(2000){Lewis}, {Challinor}, \& {Lasenby}}]{2000Lewis}
{Lewis}, A., {Challinor}, A., \& {Lasenby}, A. 2000, \apj, 538, 473

\bibitem[{{Lidz} {et~al.}(2008){Lidz}, {Zahn}, {McQuinn}, {Zaldarriaga}, \&
  {Hernquist}}]{lidz2008}
{Lidz}, A., {Zahn}, O., {McQuinn}, M., {Zaldarriaga}, M., \& {Hernquist}, L.
  2008, \apj, 680, 962

\bibitem[{{Liu} {et~al.}(2010){Liu}, {Tegmark}, {Morrison}, {Lutomirski}, \&
  {Zaldarriaga}}]{liu10}
{Liu}, A., {Tegmark}, M., {Morrison}, S., {Lutomirski}, A., \& {Zaldarriaga},
  M. 2010, \mnras, 408, 1029

\bibitem[{{Lukic} {et~al.}(2007){Lukic}, {Heitmann}, {Habib}, {Bashinsky}, \&
  {Ricker}}]{2007LukicHHBR}
{Lukic}, Z., {Heitmann}, K., {Habib}, S., {Bashinsky}, S., \& {Ricker}, P.~M.
  2007, ArXiv Astrophysics e-prints

\bibitem[{{Madau} {et~al.}(1996){Madau}, {Ferguson}, {Dickinson}, {Giavalisco},
  {Steidel}, \& {Fruchter}}]{madau1996}
{Madau}, P., {Ferguson}, H.~C., {Dickinson}, M.~E., {et~al.} 1996, \mnras, 283,
  1388

\bibitem[{{Mao} {et~al.}(2008){Mao}, {Tegmark}, {McQuinn}, {Zaldarriaga}, \&
  {Zahn}}]{mao08}
{Mao}, Y., {Tegmark}, M., {McQuinn}, M., {Zaldarriaga}, M., \& {Zahn}, O. 2008,
  ArXiv e-prints, 802

\bibitem[{{Matejek} \& {Morales}(2009)}]{matejek09}
{Matejek}, M.~S. \& {Morales}, M.~F. 2009, ArXiv e-prints

\bibitem[{{McMahon} {et~al.}(2008){McMahon}, {Parry}, {Venemans}, {King},
  {Ryan-Weber}, {Bland-Hawthorn}, \& {Horton}}]{2008McMahon}
{McMahon}, R., {Parry}, I., {Venemans}, B., {et~al.} 2008, The Messenger, 131,
  11

\bibitem[{{McQuinn} {et~al.}(2006){McQuinn}, {Zahn}, {Zaldarriaga},
  {Hernquist}, \& {Furlanetto}}]{mcquinn06}
{McQuinn}, M., {Zahn}, O., {Zaldarriaga}, M., {Hernquist}, L., \& {Furlanetto},
  S.~R. 2006, \apj, 653, 815

\bibitem[{{Mesinger} {et~al.}(2010){Mesinger}, {Furlanetto}, \&
  {Cen}}]{mesinger10}
{Mesinger}, A., {Furlanetto}, S., \& {Cen}, R. 2010, ArXiv e-prints

\bibitem[{{Mo} \& {White}(1996)}]{mo96}
{Mo}, H.~J. \& {White}, S.~D.~M. 1996, \mnras, 282, 347

\bibitem[{{Morales} {et~al.}(2006){Morales}, {Bowman}, \& {Hewitt}}]{morales06}
{Morales}, M.~F., {Bowman}, J.~D., \& {Hewitt}, J.~N. 2006, \apj, 648, 767

\bibitem[{{Nilsson} {et~al.}(2007){Nilsson}, {Orsi}, {Lacey}, {Baugh}, \&
  {Thommes}}]{2007Nilsson}
{Nilsson}, K.~K., {Orsi}, A., {Lacey}, C.~G., {Baugh}, C.~M., \& {Thommes}, E.
  2007, \aap, 474, 385

\bibitem[{{Ouchi} {et~al.}(2008){Ouchi}, {Shimasaku}, {Akiyama}, {Simpson},
  {Saito}, {Ueda}, {Furusawa}, {Sekiguchi}, {Yamada}, {Kodama}, {Kashikawa},
  {Okamura}, {Iye}, {Takata}, {Yoshida}, \& {Yoshida}}]{2008Ouchi}
{Ouchi}, M., {Shimasaku}, K., {Akiyama}, M., {et~al.} 2008, \apjs, 176, 301

\bibitem[{{Pritchard} \& {Furlanetto}(2006)}]{pritchard2006}
{Pritchard}, J.~R. \& {Furlanetto}, S.~R. 2006, \mnras, 367, 1057

\bibitem[{{Pritchard} \& {Furlanetto}(2007)}]{pritchard06}
{Pritchard}, J.~R. \& {Furlanetto}, S.~R. 2007, \mnras, 376, 1680

\bibitem[{{Pritchard} \& {Loeb}(2008)}]{pritchard2008}
{Pritchard}, J.~R. \& {Loeb}, A. 2008, \prd, 78, 103511

\bibitem[{{Pritchard} \& {Loeb}(2010)}]{pritchard10}
{Pritchard}, J.~R. \& {Loeb}, A. 2010, \prd, 82, 023006

\bibitem[{{Reed} {et~al.}(2007){Reed}, {Bower}, {Frenk}, {Jenkins}, \&
  {Theuns}}]{2007ReedBFJT}
{Reed}, D.~S., {Bower}, R., {Frenk}, C.~S., {Jenkins}, A., \& {Theuns}, T.
  2007, \mnras, 374, 2

\bibitem[{{Salvaterra} {et~al.}(2010){Salvaterra}, {Ferrara}, \&
  {Dayal}}]{salvaterra10}
{Salvaterra}, R., {Ferrara}, A., \& {Dayal}, P. 2010, ArXiv e-prints

\bibitem[{{Santos} {et~al.}(2008){Santos}, {Amblard}, {Pritchard}, {Trac},
  {Cen}, \& {Cooray}}]{santos08}
{Santos}, M.~G., {Amblard}, A., {Pritchard}, J., {et~al.} 2008, \apj, 689, 1

\bibitem[{{Santos} {et~al.}(2005){Santos}, {Cooray}, \& {Knox}}]{santos05}
{Santos}, M.~G., {Cooray}, A., \& {Knox}, L. 2005, \apj, 625, 575

\bibitem[{{Santos} {et~al.}(2010){Santos}, {Ferramacho}, {Silva}, {Amblard}, \&
  {Cooray}}]{santos10}
{Santos}, M.~G., {Ferramacho}, L., {Silva}, M.~B., {Amblard}, A., \& {Cooray},
  A. 2010, \mnras, 406, 2421

\bibitem[{{Schaerer}(2003{\natexlab{a}})}]{2003Schaerer}
{Schaerer}, D. 2003{\natexlab{a}}, \aap, 397, 527

\bibitem[{{Schaerer}(2003{\natexlab{b}})}]{schaerer03}
{Schaerer}, D. 2003{\natexlab{b}}, \aap, 397, 527

\bibitem[{{Schilizzi} {et~al.}(2007){Schilizzi}, {Alexander}, {Cordes},
  {Dewdney}, {Ekers}, \& {Faulkner}}]{schilizzi07}
{Schilizzi}, R.~T., {Alexander}, P., {Cordes}, J.~M., {et~al.} 2007, SKA
  documents: http://www.skatelescope.org

\bibitem[{{Semelin} {et~al.}(2007){Semelin}, {Combes}, \& {Baek}}]{semelin2007}
{Semelin}, B., {Combes}, F., \& {Baek}, S. 2007, \aap, 474, 365

\bibitem[{{Shaw} \& {Lewis}(2008)}]{shaw2008}
{Shaw}, J.~R. \& {Lewis}, A. 2008, \prd, 78, 103512

\bibitem[{{Stark} {et~al.}(2007){Stark}, {Ellis}, {Richard}, {Kneib}, {Smith},
  \& {Santos}}]{2007Stark}
{Stark}, D.~P., {Ellis}, R.~S., {Richard}, J., {et~al.} 2007, \apj, 663, 10

\bibitem[{{Stiavelli} {et~al.}(2004){Stiavelli}, {Fall}, \&
  {Panagia}}]{2004Stiavelli}
{Stiavelli}, M., {Fall}, S.~M., \& {Panagia}, N. 2004, \apjl, 610, L1

\bibitem[{{Trac} \& {Cen}(2007)}]{2007Trac}
{Trac}, H. \& {Cen}, R. 2007, \apj, 671, 1

\bibitem[{{Trac} {et~al.}(2008){Trac}, {Cen}, \& {Loeb}}]{trac08}
{Trac}, H., {Cen}, R., \& {Loeb}, A. 2008, \apjl, 689, L81

\bibitem[{{Trac} \& {Pen}(2004)}]{2004TracPen}
{Trac}, H. \& {Pen}, U.-L. 2004, New Astronomy, 9, 443

\bibitem[{{Trac} \& {Pen}(2006)}]{2006TracPen}
{Trac}, H. \& {Pen}, U.-L. 2006, New Astronomy, 11, 273

\bibitem[{{Willis} {et~al.}(2008){Willis}, {Courbin}, {Kneib}, \&
  {Minniti}}]{2008Willis}
{Willis}, J.~P., {Courbin}, F., {Kneib}, J., \& {Minniti}, D. 2008, \mnras,
  384, 1039

\bibitem[{{Wouthuysen}(1952)}]{wouth1952}
{Wouthuysen}, S.~A. 1952, \aj, 57, 31

\bibitem[{{Zahn} {et~al.}(2007){Zahn}, {Lidz}, {McQuinn}, {Dutta}, {Hernquist},
  {Zaldarriaga}, \& {Furlanetto}}]{zahn07}
{Zahn}, O., {Lidz}, A., {McQuinn}, M., {et~al.} 2007, \apj, 654, 12

\end{thebibliography}

\end{document}